\documentclass[preprint,12pt,authoryear]{agujournal}

\usepackage{graphicx}
\usepackage{rotating}
\usepackage{bm}
\usepackage{color}
\usepackage[titletoc,title]{appendix}

\usepackage{amssymb}
\usepackage{amsmath}
\usepackage{longtable}
\usepackage{array}
\usepackage{natbib}
\newcolumntype{L}{>$1<$}

\journalname{Journal of Advances in Modeling Earth Systems (JAMES)}

\begin{document}

\title{The KPP boundary layer scheme: revisiting its formulation and benchmarking one-dimensional ocean simulations relative to LES}


\authors{Luke Van Roekel\affil{1}, Alistair J. Adcroft\affil{2}, Gokhan Danabasoglu\affil{3}, Stephen M. Griffies\affil{2}, Brian Kauffman\affil{3}, William Large\affil{3}, Michael Levy\affil{3}, Brandon Reichl\affil{2}, Todd Ringler\affil{1}, Martin Schmidt\affil{4}}

\affiliation{1}{Fluid Dynamics and Solid Mechanics, Los Alamos National Laboratory, Los Alamos, NM, LA-UR-16-26992}
\affiliation{2}{NOAA / Geophysical Fluid Dynamics Laboratory, Princeton, NJ and Princeton University Program in Atmospheric and Oceanic Sciences, Princeton, NJ}
\affiliation{3}{Climate and Global Dynamics Laboratory, National Center for Atmospheric Research, Boulder, CO}
\affiliation{4}{Leibniz-Institute of Baltic Sea Research, Warnem\"unde, Seestra\ss e 15, 18119, Rostock, Germany}

\correspondingauthor{Luke Van Roekel}{lvanroekel@lanl.gov}

\begin{keypoints}
\item  The Community ocean Vertical Mixing (CVMix) project version of the K-profile parameterization (KPP) is compared across a suite of oceanographically relevant regimes against large eddy simulations (LES).
\item The standard configuration of KPP is consistent with LES results across many simulations, but some adaptations of KPP for improved applicability relative to LES comparisons are proposed.
\item An alternate, computationally simpler, configuration of KPP is proposed to alleviate the need to represent rapidly changing diffusivities near the base of the ocean surface boundary layer.
\end{keypoints}
\begin{abstract}

We evaluate the Community ocean Vertical Mixing (CVMix) project version of the K-profile parameterization (KPP). For this purpose, one-dimensional KPP simulations are compared across a suite of oceanographically relevant regimes against large eddy simulations (LES).  The LES is forced with horizontally uniform boundary fluxes and has horizontally uniform initial conditions, allowing its horizontal average to be compared to one-dimensional KPP tests.  We find the standard configuration of KPP \citep{Danabasoglu2006} consistent with LES across many forcing regimes, supporting the physical basis of KPP. Our evaluation motivates recommendations for "best practices" for using KPP within ocean circulation models, and identifies areas where further research is warranted. Further, our test suite can be used as a baseline for evaluation of a broad suite of boundary layer models.

The original treatment of KPP recommends the matching of interior diffusivities and their gradients to the KPP predicted values computed in the ocean surface boundary layer (OSBL). However, we find that difficulties in representing  derivatives of rapidly changing diffusivities near the base of the OSBL can lead to loss of simulation fidelity.  We propose two alternative approaches: (1) match to the internal predicted diffusivity along, (2) set the KPP diffusivity to zero at the OSBL base. Although computationally simpler, the second alternative is sensitive to implementation details and we offer methods to prevent the emergence of numerical high frequency noise.  

We find the KPP entrainment buoyancy flux to be sensitive to vertical grid resolution and details of how to diagnose the KPP boundary layer depth. We modify the KPP turbulent shear velocity parameterization to reduce resolution dependence. Additionally, our results show that the KPP parameterized non-local tracer flux is incomplete due to the assumption that it solely redistributes the surface tracer flux.  However, examination of the LES vertical turbulent scalar flux budgets show that non-local fluxes can exist in the absence of surface tracer fluxes.  This result motivates further studies of the non-local flux parameterization.  
\end{abstract}

\begin{center} {\bf Draft from} \today  \end{center}


\section{Introduction}
\label{Intro_Mixing_background}

The ocean surface boundary layer (OSBL) mediates momentum, heat, and scalar tracer fluxes between the interior ocean with the atmosphere and cryosphere. Consequently, an accurate parameterization of turbulence and the induced vertical mixing in the OSBL is essential for robust model simulations of climate physics and of the abundance and distribution of important biological and chemical quantities.  We here consider the formulation and behavior of the K-profile parameterization (KPP) (\citealp{Large1994}, LMD94), which has been used by a variety of ocean and climate applications.  Here we consider one-dimensional vertical fluxes only.  KPP fidelity in the presence of horizontal features (\textit{e.g.}, baroclinic fronts; \citealp{bachman2017parameterization}) is not considered.

\subsection{The suite of boundary layer parameterizations}

Generally, boundary layer parameterizations like KPP assume the turbulent mixing is dominated by vertical fluxes. Presently, varying degrees of complexity are used to parameterize these fluxes.  Bulk boundary layer models are perhaps the simplest (\textit{e.g.}, \citealp{kraus1967one,Niller1977one,Price1986,gaspar1988modeling}), where ocean properties (tracers and momentum) are assumed to be vertically uniform in the OSBL.  The assumption of no vertical structure within the boundary layer is both the key simplification and the main deficiency of bulk models.  Namely, ocean tracers and momentum are not generally uniform vertically, even in the presence of strong mixing.  Hence, vertical integration over the depth of the OSBL precludes the simulation of OSBL processes such as the Ekman spiral and boundary layer restratification \citep[but see][for a proposed method to remedy these deficiencies]{Hallberg_mixed}. 

Turbulence Kinetic Energy (TKE) closure (TC) is another widely used framework for parameterizing upper ocean boundary layer mixing (\textit{e.g.}, \citealp{kantha1994improved,Canuto2001,Umlauf2005,canuto2007non,harcourt2015improved}). In most variants of TC, the profiles of eddy diffusivity and viscosity are dependent on the local TKE, which is prognostic (\textit{e.g.}, \citealp{Mellor1982,kantha1994improved}).  Most of these models are local as the turbulent fluxes of tracers and momentum exist only in the presence of non-zero vertical gradients of the mean quantities. Yet in highly convective conditions strong fluxes exist even when $\partial \theta / \partial z \approx 0$.  Thus many TC models are deficient in highly convective conditions, as they predict zero flux through much of the boundary layer. However, a few TC models (\textit{e.g.}, \citealp{stull1993review,lappen2001toward,soares2004eddy}) do include non-local mixing effects.

The K-profile parameterization (KPP) (\citealp{Large1994}, hereafter LMD94) aims to represent a middle ground between bulk boundary layer models and prognostic TC models.  KPP allows for vertical property variations in the OSBL via a specified vertical shape function \citep{OBrien1970}.  KPP includes a parameterized non-local transport allowing for the existence of vertical turbulent scalar fluxes in the absence of vertical gradients of scalar quantities.

\begin{figure}[thbp]
	\centering
    \includegraphics[width=1\linewidth]{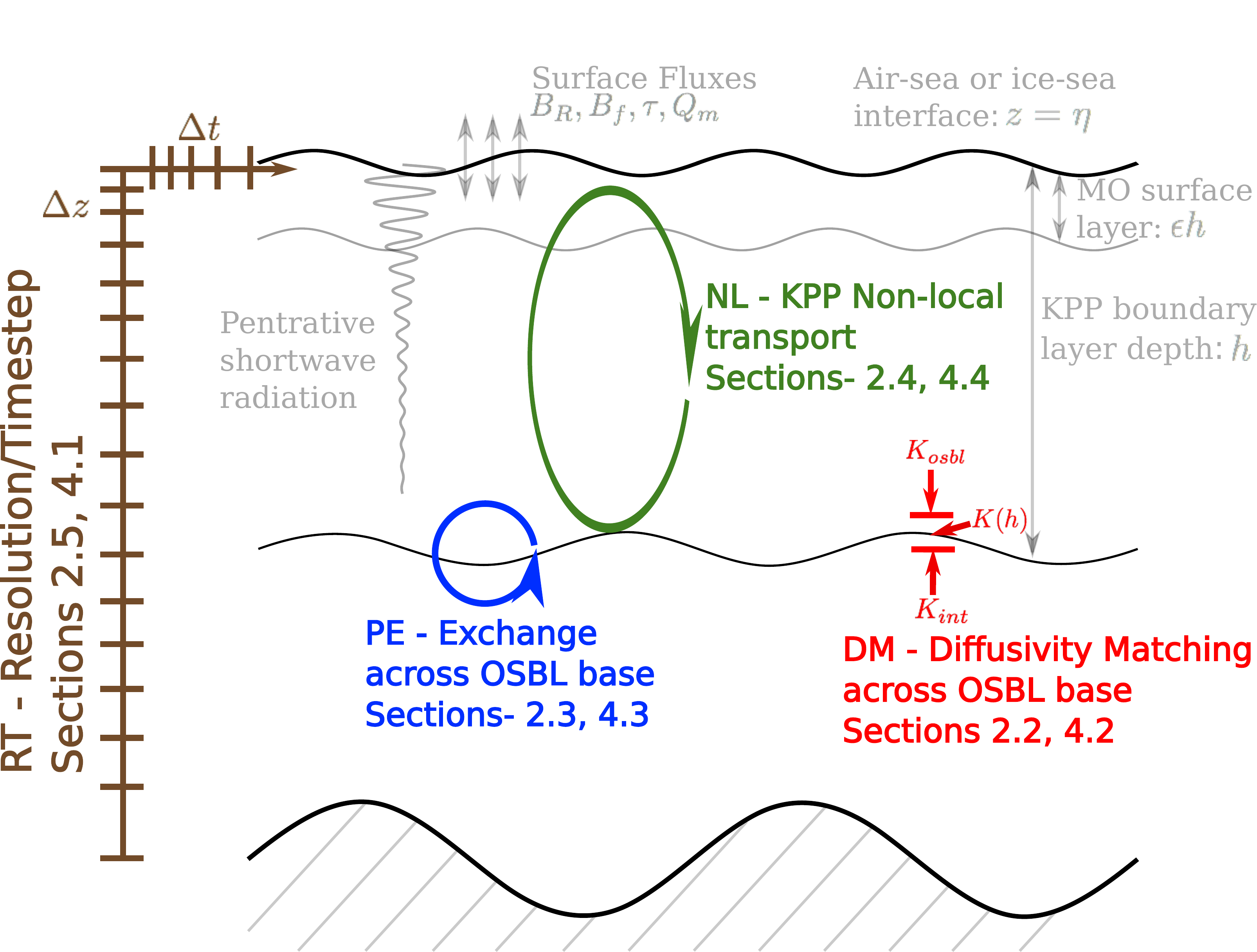}
\caption{Schematic of the upper ocean regions associated with the KPP boundary layer parameterization.  The upper ocean is exposed to fluxes of momentum, $\tau$, mass, $Q_m$, and buoyancy, $B_f$, at the air-sea interface.  Penetrating shortwave radiation, $(\overline{w^\prime \theta^\prime})_R$, along with its associated buoyancy flux, $B_{R}>0$, also influence the upper ocean.  The Monin-Obukhov (MO) surface layer transfers these fluxes to the remainder of the boundary layer.  Here we assume the surface layer extends from the free surface ($z = \eta$) to $z = \eta - \epsilon h$, where $h > 0$ is the depth of the boundary layer base (typically $\epsilon \approx 0.1$).  The colored portions of the schematic represent the four components of KPP we focus on in this paper (Section~\ref{issues}).  The red section represents the diffusivity matching in KPP, where $K_{osbl}$, $K_{int}$, and $K(h)$ are the diffusivities at the last layer in the OSBL, the first model layer below the OSBL, and at the OSBL respectively.  The green signifies the non-local parameterization, the blue represents entrainment and detrainment across the OSBL base, and the brown represents the influence of choices in the calling model on KPP fidelity.  The sections for each color denotes the portions of the paper where that process is examined.}
\label{BL_schematics}
\end{figure}

\subsection{The CVMix Project}

KPP has been implemented in numerous ocean circulation models. In our experience, each implementation makes slightly distinct physical and numerical choices. Sometimes, these implementation choices have unintended consequences that can negatively impact the KPP boundary layer simulation.  This situation provided the mandate for our development of KPP within the CVMix project \citep{Griffies2015}, and the examination of that implementation in this paper. 

The CVMix library is developed as a suite of standardized vertical mixing parameterizations to be implemented in a three-dimensional ocean circulation model.  As a part of the CVMix development, our purpose here is to evaluate the influence of physical assumptions and numerical choices within the CVMix version of KPP\footnote{For brevity, in this paper ``KPP'' refers to its CVMix implementation, which is fully consistent with the description in \cite{Danabasoglu2006}. The Langmuir Turbulence parameterization of \cite{Li:2015gv} is also implemented, but not evaluated in this work.} as well as in the ocean circulation model.  The present paper serves to document many of our findings.  

\subsection{An LES evaluation framework for OSBL parameterizations}

Large Eddy Simulations (LES) provide our evaluation framework for the KPP scheme. As configured here, the LES resolves the dominant eddies and with sub-grid turbulence small relative to that realized at resolved scales. The LES is forced with horizontally uniform surface fluxes and the output is horizontally and temporally averaged. We compare the LES results to one-dimensional KPP simulations using identical forcing and initialization. Numerous studies have compared atmospheric boundary layer parameterizations to LES \citep[\textit{e.g.},][]{Holtslag1991,Moeng1994,Brown1996,Ayotte1996,Noh2003}, with far fewer LES evaluations of ocean parameterizations.  Those ocean LES comparisons have made similar (and limited) initial conditions and forcing scenarios \citep[\textit{e.g.},][]{McWilliams2000,Smyth2002,Reichl:2016db,bachman2017parameterization}. Our somewhat larger test suite facilitates an examination of the KPP under a broader range of forcing scenarios. 

Following atmospheric comparisons \citep[\textit{e.g.},][]{Ayotte1996}, we use a suite of forcing  and initial conditions. However, our LES configuration precludes the examination of interactions with horizontal processes in the boundary layer, such as from submesoscale frontal instabilities \citep[\textit{e.g.},][]{boccaletti_etal2007,foxkemper_etal2008,bachman2017parameterization}. Instead, we focus on examining the influence of physical assumptions within KPP as well as numerical implementation choices. 

We test the CVMix implementation of KPP against LES configured with similar depth and surface forcing. Furthermore, simulation differences can arise due to details of how KPP is embedded within the calling model.  We thus compare KPP as realized in three ocean circulation models: Model for Prediction Across Scales - Ocean (MPAS-O; \citep{ringler2013multi}); Modular Ocean Model Version 6 (MOM6;  \citep{AdcroftHallberg2006}); and Parallel Ocean Program Version 2 (POP2; \citep{Smith_POP_2010}). We found these details explain some differences in behavior relative to LMD94, and thus they are important for evaluating the integrity of KPP and for choosing its parameters. 

\subsection{Organization of this paper}

We start the main portion of this paper in Section~\ref{issues},  where we summarize elements of KPP and present salient implementation considerations and issues that arose in our development and evaluation. We then present a few sections that document our experiences benchmarking KPP against LES.  We discuss the test cases and LES model in Section \ref{test_cases}, and then describe results and analysis in Sections \ref{section:RTresults}-\ref{section:NLresults}.  The presentation of results focuses on issues discussed in Section \ref{issues} and offers possible solutions. We close the main portion of the paper with conclusions and best practice recommendations in Section \ref{discussion}, and then with suggested future KPP developments in Section \ref{section:futureWork}.  Various appendices offer further details about aspects of the KPP scheme and the test cases.

\section{KPP considerations}
\label{issues}
The implementation of KPP within CVMix closely follows that detailed in LMD94, including the modifications of \cite{Danabasoglu2006}.  We present here some salient points that elucidate implementation considerations and introduce issues that arose during our CVMix development. Further formulation details of the KPP scheme are given in Appendix~\ref{sec:KPP-describe}, and a Table of symbols along with preferred units is given in Appendix~\ref{math_symbols}. 

\subsection{Basic definitions}

The vertical position in the ocean, $z$, ranges from 
\begin{equation}
-H(x,y) \le z \le \eta(x,y,t)
\end{equation}
where $z=-H$ is the position of the static ocean bottom, $z=\eta$ is the position for the dynamic ocean free surface, $z=0$ is the resting ocean surface (reference geopotential), and
 \begin{equation}
  \mbox{depth} = -z + \eta \ge 0 
 \end{equation}
is the positive distance from the ocean surface to a point in the ocean. When referring to positions within the surface boundary layer, it is convenient to use a non-dimensional boundary layer coordinate
\begin{equation}
\label{sigDefn}
\sigma \equiv \frac{-z + \eta}{h} 
\qquad 0 \le \sigma \le 1,
\end{equation}
where $\sigma=0$ at the ocean free surface and $\sigma=1$ at the boundary layer base where 
\begin{equation}
 -z + \eta = h  \qquad \mbox{OSBL thickness}.
\label{eq:osbl-base-thickness}
\end{equation}

\subsection{General structure of the KPP parameterization}

For any prognostic scalar or vector field component $\psi$ (\textit{e.g.}, velocity components, tracer concentrations), the KPP scheme parameterizes the turbulent vertical flux within the surface boundary layer according to 
\begin{equation}
\label{KPPfluxText}
\overline{w^\prime \psi^\prime} = -K_\psi \frac{\partial \psi}{\partial z} + K_\psi \gamma_\psi.
\end{equation}
In this equation, the first term represents the local contribution to the turbulent vertical flux of $\psi$, and the second term is the parameterized non-local flux.  The eddy diffusivity $K_\psi$ is written as the product of three terms
\begin{equation}
\label{KPPDiffusivityText}
K_\psi = h\, w_\psi \left(\sigma\right)G\left(\sigma \right),
\end{equation}
where $h$ is the KPP diagnosed OSBL depth (equation (\ref{eq:osbl-base-thickness}), and $G(\sigma)$ is the vertical shape function.

Equations~(\ref{KPPfluxText})-(\ref{KPPDiffusivityText}) suggest a few important considerations in the implementation of KPP. For example, the boundary layer depth $h >0$ scales the diffusivity, so that the maximum $K_\psi$ is larger for deeper boundary layers, suggesting that KPP is critically dependent on the boundary layer depth.  Further, the vertical structure of the KPP diffusivity and hence the non-local flux is set by the non-dimensional shape function, $G(\sigma)$, which is scaled by the boundary layer depth.  

In this work we focus on four main physical and numerical processes colored in Figure~\ref{BL_schematics}:
\begin{itemize}
 \item matching the diffusivity and its gradient (DM) at the OSBL base (red);
  \item property exchange (PE; \textit{i.e.}, entrainment and detrainment) across the OSBL base (blue);
  \item non-local (NL) transport (green);
  \item model resolution and timestep (RT; brown).
\end{itemize}
Given that all of the processes in Figure~\ref{BL_schematics} are influenced by the KPP-diagnosed OSBL depth, the KPP OSBL depth algorithm is pivotal and discussed in some detail in Section~\ref{blDepthAlgorithm}. Our four focus areas are introduced in the following subsections.

\subsection{Bulk Richardson number and the boundary layer depth}
\label{blDepthAlgorithm}

In KPP, the bulk Richardson number is computed by
\begin{equation}
\mbox{Ri}_{\mbox{\tiny b}} = \frac{ (b_{\tiny \mbox{sl}} - b(z)) \, (-z+\eta) }{|\mathbf{u}_{\tiny \mbox{sl}} - \mathbf{u}(z)|^2 + V_t^2(z) }.
\label{eq:rib-defined}
\end{equation}
In this expression, $b$ is the buoyancy (dimensions of length per squared time) based on the surface referenced potential density, and $b_{\tiny \mbox{sl}}$ is the  buoyancy averaged over the depth of the surface layer between the depth range $0 \le \sigma \le \epsilon$ (see Figure \ref{BL_schematics}). Small differences of $b_{\tiny \mbox{sl}} - b(z)$ signal weak vertical stratification, characteristic of a region within the surface boundary layer. In contrast, large differences arise when $z$ reaches into the more stratified region beneath the boundary layer. The denominator in $\mbox{Ri}_{\mbox{\tiny b}}$ consists of the squared vertical shear resolved by the model's prognostic horizontal velocity field:   
$|\mathbf{u}_{\tiny \mbox{sl}} - \mathbf{u}(z)|^2$, where $\mathbf{u}_{\tiny \mbox{sl}}$ is the surface layer averaged horizontal velocity. In addition, the term $V_t^2$ aims to parameterize unresolved vertical shears near the OSBL base (see Section~\ref{section:vt2}). When either the resolved or parameterized shear is large, the bulk Richardson number is small and the OSBL deepens.  Finally, note that even if the buoyancy difference and vertical shear are vertically constant, the bulk Richardson number increases linearly with depth, $d = -z+\eta$, given the presence of depth in the numerator of equation~(\ref{eq:rib-defined}).

The depth at the base of the ocean surface boundary layer, $-z+\eta = h$, is computed as the depth where the bulk Richardson number equals a critical value
\begin{equation}
\mbox{Ri}_{\mbox{\tiny b}} = \mbox{Ri}_{\mbox{\tiny crit}} = \frac{ (b_{\tiny \mbox{\tiny \mbox{sl}}} - b(h)) \, h }{|\mathbf{u}_{\tiny \mbox{\tiny \mbox{sl}}} - \mathbf{u}(h)|^2 + V_t^2(h) }.
\label{eq:KPP-boundary-layer-depth-defined}
\end{equation}
In LMD94, the critical bulk Richardson number was set to $\mbox{Ri}_{\mbox{\tiny crit}} = 0.3$. However, values between 0.25 and 1.0 have been used in similar formulations (\textit{e.g.}, \citealp{troen1986simple,Vogelezang1996,Mcgrath-Spangler2015}). In general, the correct diagnosis of the boundary layer depth is a key part of the KPP scheme, as this depth controls the upper ocean turbulent layer and the strength of mixing within that layer.  The diagnosed boundary layer depth also controls the strength of entrainment into the OSBL.  We thus expect the chosen definitions of surface layer fraction ($\epsilon$; Figure~\ref{BL_schematics}),  parameterized turbulent vertical shear, $V_t^2$, and the critical bulk Richardson number, $\mbox{Ri}_{\mbox{\tiny crit}}$, to strongly influence KPP results and their comparison to LES.

Values of $\mbox{Ri}_{\mbox{\tiny crit}}$ vary from $\mbox{Ri}_{\mbox{\tiny crit}} = 0.25$ for shear instability when using a linear stability analysis \citep{miles1961stability}\footnote{This definition may not be appropriate for weak mean shear and breaking internal waves \citep{troy2005,barad2010}.} to O(1) for a non-linear stability analysis \citep{abarbanel1984richardson}.  However, \cite{troen1986simple} argue that shear may not be adequately resolved in a model simulation, prompting use of a larger value of $\mbox{Ri}_{\mbox{\tiny crit}}$ that is a function of vertical grid spacing. 


Further, it is  unlikely that the bulk Richardson number computed at a model interface is exactly equal to the chosen critical threshold. Consequently, the KPP boundary layer depth, $h$, will also be sensitive to the interpolation method used to determine where $\mbox{Ri}_{\mbox{\tiny b}} = \mbox{Ri}_{\mbox{\tiny crit}}$ according to equation~(\ref{eq:KPP-boundary-layer-depth-defined}) \citep{Danabasoglu2006,Seidel2010}. For direct comparison between KPP and LES we use the KPP bulk Richardson number method based on equation~(\ref{eq:KPP-boundary-layer-depth-defined}) to determine the OSBL depth with LES data. 

\subsection{Shape function and diffusivity matching}
\label{sec:shapefunc-diffusmatch}
 
The vertical shape function ($G(\sigma)$) controls the vertical structure of diffusivity and the non-local flux in KPP (equation (\ref{KPPDiffusivityText})).  The shape function is assumed to follow a cubic polynomial as proposed by \cite{OBrien1970}.  The method to determine all of the coefficients of the polynomial is given in Appendix \ref{subsection:non-dimensional-shape-function} and LMD94.  Here we identify potential issues with the matching of diffusivities from KPP to those determined by interior mixing parameterizations.

\subsubsection{Problems matching diffusivity derivatives at the OSBL base}

The KPP diffusivity matching proposed by LMD94 means that parameterized turbulence generated in the ocean interior (\textit{e.g.}, internal waves, shear instabilities) indirectly influences that in the surface boundary layer.  Given that the interior diffusivity can have nontrivial vertical structure, its vertical gradient can be large and potentially negative.  Hence, there is no guarantee that the KPP diffusivity will vary smoothly across the OSBL base.  Further, the vertical derivative of the diffusivity can be sharp (\textit{i.e.}, discontinuous) near the OSBL base, which leads to a generally poor representation on a discrete vertical grid.  Therefore, given these potential difficulties we test the sensitivity of KPP to three variants of internal matching: matching the internal diffusivity and its gradient, matching the internal diffusivity alone, and abandoning internal matching in section~\ref{section:DMresults}. 

\subsubsection{Problems with the non-local parameterization at the OSBL base}

If the KPP computed boundary layer diffusivities (and the derivatives) are forced to match the corresponding interior values, the vertical shape function, $G(\sigma)$, is not guaranteed to be zero at the base of the boundary layer, $G(
\sigma=1) \ne 0$.  A non-zero $G(1)$ is not desirable since it leads to a non-local flux at the boundary layer base. The non-local flux should vanish at the base of the boundary layer since that is the depth at which boundary layer turbulence vanishes. 

Neglecting diffusivity matching at the OSBL base would resolve this inconsistency for the non-local term.  As a physical justification for dropping the diffusivity matching, one may assume that KPP is only parameterizing effects from the larger and stronger turbulent boundary layer eddies.  Other interior mixing parameterizations (\textit{e.g}, the LMD94 shear instability mixing scheme) could be assumed to be scale separated from KPP as KPP seeks to parameterize penetration of the most unstable eddies.  Under this assumption, the net diffusivity in the OSBL is computed as the sum of that predicted by KPP plus other parameterizations.  In Section~\ref{section:DMresults}, we examine a configuration of KPP where diffusivity matching is neglected.

\subsection{Property exchanges across the OSBL base}
\label{section:entrainment}

Property exchanges at the OSBL base are determined in two ways: a parameterization of unresolved turbulence ($V_t^2$) and the enhanced diffusivity parameterization.  The former exerts a strong influence at high vertical resolution, while the latter dominates at coarser vertical resolution.  We discuss both in this section.

\subsubsection{The parameterization of $V_t^2$ in $\mbox{Ri}_{\mbox{\tiny b}}$}
\label{section:vt2}

\begin{figure}[thbp]
\centering\includegraphics[width=0.85\linewidth]{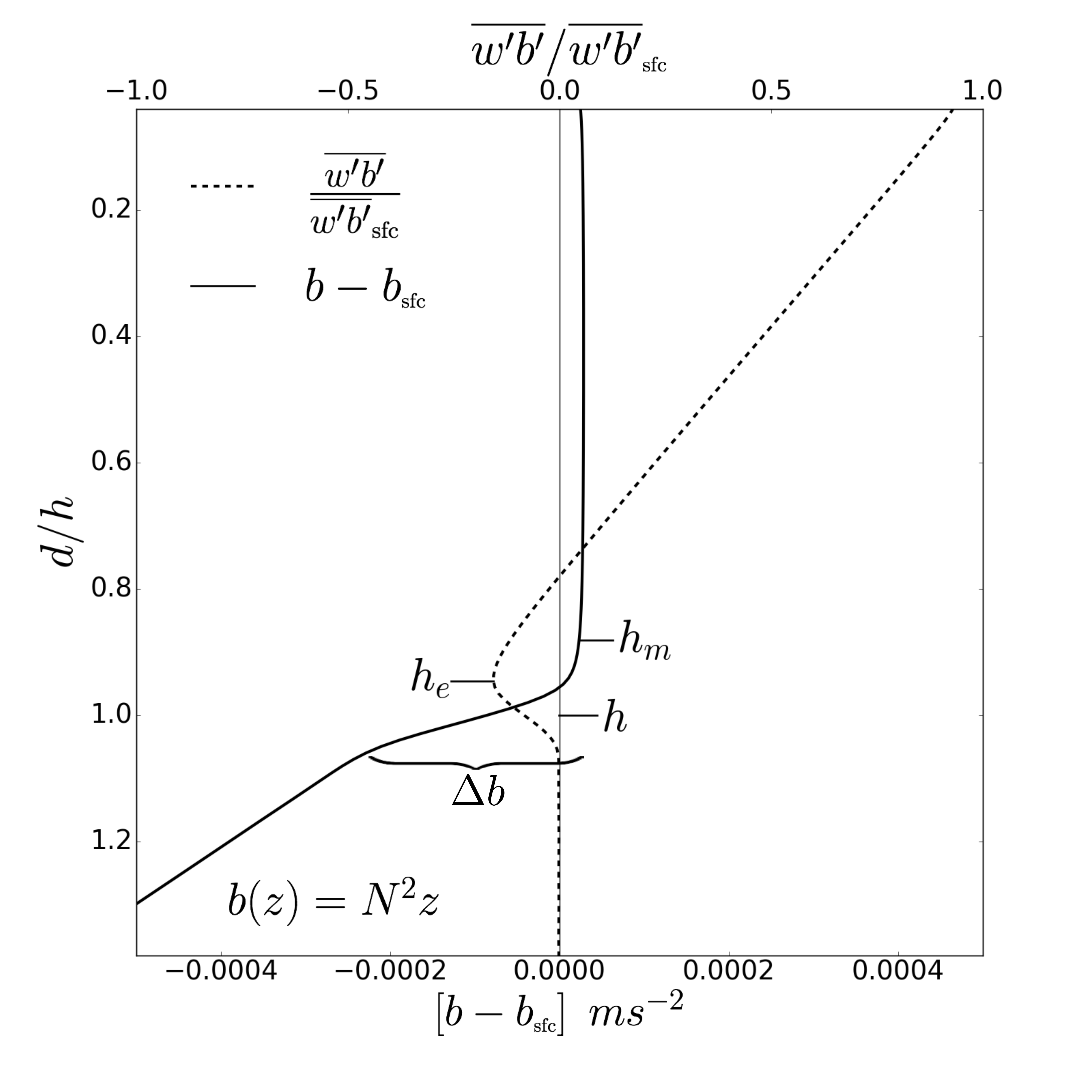}
\caption{Schematic illustrating the various depths arising in the boundary layer as forced by a destabilizing surface buoyancy flux $\overline{w^\prime b^\prime}_{\mbox{\tiny sfc}} < 0$ (defined in equation~(\ref{eq:wb})). This forcing supports active non-local convective boundary layer turbulence. Data for this figure is taken from the LES results of the FC experiment (Figure~\ref{matchconfig_compare}a). The vertical axis is the non-dimensional boundary layer depth, $\sigma = (-z+\eta)/h$, extending from beneath the boundary layer ($\sigma > 1$) to the base of the surface layer ($\sigma = \epsilon \, h$) (see Figure \ref{BL_schematics}). The dashed line is the local turbulent buoyancy flux, $(\overline{w^\prime b^\prime})$,  normalized by the surface buoyancy flux $(\overline{w^\prime b^\prime}_{\mbox{\tiny sfc}})$. The scale for this ratio is along the top axis. Depths where $\overline{w^\prime b^\prime}/\overline{w^\prime b^\prime}_{\mbox{\tiny sfc}} < 0$ are where the local turbulence stabilizes the boundary layer, which occurs near the boundary layer base. The solid line is the difference in local buoyancy and  the surface layer buoyancy, $b - b_{\mbox{\tiny sfc}}$, with corresponding scale along the bottom axis (in units of $10^{-4}$ $\mbox{m}~\mbox{s}^{-2}$). The mixed layer is  where buoyancy is weakly stratified,  
$\partial b/\partial z \approx 0$. The mixed layer base is determined subjectively by $\partial b/\partial z > (\partial b/\partial z)_{\mbox{\tiny min}}> 0$, with $(\partial b/\partial z)_{\mbox{\tiny min}}$ a chosen minimum stratification criteria. For much of the boundary layer, $b - b_{\mbox{\tiny sfc}} > 0$, since the surface layer buoyancy is driven low by the destabilizing surface flux. The entrainment depth, $h_e$, is where $\overline{w^\prime b^\prime}/\overline{w^\prime b^\prime}_{\mbox{\tiny sfc}} < 0$ reaches a minimum, with $h_e$ straightforward to diagnose in a LES. (See \ref{analytic} for an idealized buoyancy profile similar to that shown here allowing for an analytic expression for $h$). In the entrainment layer, buoyancy changes rapidly ($\Delta b$), reflecting enhanced vertical stratification below the mixed layer.   Below the entrainment layer, the buoyancy profile is roughly constant ($b \approx N^2 \, z$ with $N^{2} > 0$ constant and $z<0$). The KPP boundary layer depth, $h$, is determined by the bulk Richardson number criteria in equation~(\ref{eq:KPP-boundary-layer-depth-defined}).}
\label{Vt_schematics}
\end{figure}

The KPP scheme includes a term related to an energy associated with unresolved turbulence ($V_t^2$) in the bulk Richardson number denominator in equation~(\ref{eq:rib-defined}).  The purpose of this term is to sufficiently deepen the OSBL to ensure that the empirical rule of convection (see LMD94) is satisfied.  This rule says that the minimum turbulent buoyancy flux within the boundary layer is given by 
\begin{equation}
 \overline{w^\prime b^\prime}_{\mbox{\tiny $h_e$}} 
 \approx 
 -0.2 \,  
 \overline{w^\prime b^\prime}_{\mbox{\tiny sfc}},
 \label{eq:empiricalRuleConvection}
 \end{equation}
where $h_e$ is the entrainment depth (see Figure~\ref{Vt_schematics}).  The entrainment depth is where water from below the boundary layer is exchanged with boundary layer water.  The entrainment depth is also the depth of the minimum buoyancy flux.  

The parameterization of $V_t^2$ is derived by considering a buoyancy profile that is well-mixed to a given depth ($h_m$) with linear stratification below (Figure~\ref{Vt_schematics}).  The buoyancy flux at $h_e$ is written (using equations~(\ref{KPPfluxText}) and~(\ref{KPPDiffusivityText})) as
\begin{equation}
\label{buoyFluxhe}
\overline{w^\prime b^\prime}_{\mbox{\tiny $h_e$}} = -h \, w_s(\sigma) \, G(\sigma) \, 
\left(\frac{\partial b}{\partial z} - \gamma_b \right)
\qquad \mbox{with} \; \; \sigma =  \frac{h_e}{h}.
\end{equation}

Near the boundary layer depth, $h$, the non-local term $\gamma_b$ (see equation~(\ref{eq:gammab-defined})) is small and can be ignored.  Furthermore, in convective conditions, 
\begin{equation}
\label{wsapprox}
 w_b(h_e/h) 
 \rightarrow \kappa \left(c_s \, \kappa \, \epsilon\right)^{1/3} w_*  \, ,
 \end{equation}
  where $\kappa$ is the von K\'arm\'an constant, $c_s$ is an empirical constant, and $w_{*}$ is the convective turbulent velocity scale defined as
  \begin{equation}
w_* \equiv (-h \, \overline{w^\prime b^\prime}_{\mbox{\tiny sfc}} )^{1/3}.
\label{eq:convective-velocity-scale}
\end{equation}
Also, noting that $N_{osbl}^2 = (\partial b/ \partial z)_{osbl}$ and $h_e/h \approx 1$, equation~(\ref{buoyFluxhe}) becomes
\begin{eqnarray}
\overline{w^\prime b^\prime}_{h_e} &=& -\kappa \left(c_s \kappa \epsilon\right)^{1/3} w_* \, \left(1 - \frac{h_e}{h}\right)^2 \, h \, N_{osbl}^2 \nonumber \\
&=& -\kappa \left(c_s \kappa \epsilon \right)^{1/3} w_* \left(h - h_e\right)^2 N_{osbl}^2/h. 
\label{buoyFluxVt}
\end{eqnarray}
At the OSBL base, $\mbox{Ri}_{\mbox{\tiny b}} = \mbox{Ri}_{\mbox{\tiny crit}}$ (see equation~(\ref{eq:KPP-boundary-layer-depth-defined})).  We can use this result to write an equation for $V_t^2$.  By assuming the resolved vertical shear is zero we have
\begin{equation}
\label{RicVT}
\mbox{Ri}_{\mbox{\tiny crit}}= \frac{h \, (b_{\tiny \mbox{sl}} - b(h))}{V_t^2}.
\end{equation}
Given the assumed linear buoyancy profile, the numerator of equation~(\ref{RicVT}) is $h \, N_{osbl}^2 (h-h_e)$, which implies 
\begin{equation}
\label{h_he_defn}
h - h_e = \frac{V_t^2 \, \mbox{Ri}_{\mbox{\tiny crit}}}
{h \, N_{osbl}^2}.
\end{equation}
Using equations~(\ref{RicVT}) and~(\ref{h_he_defn}) brings equation~(\ref{buoyFluxVt}) into the form 
\begin{equation}
\label{interm_eqn}
\overline{w^\prime \, b^\prime}_{\mbox{\tiny $h_e$}} = 
-\kappa \left(c_s \kappa \epsilon \right)^{1/3} w_* \, V_t^4 \, \mbox{Ri}_{\mbox{\tiny crit}}^2 \, N_{osbl}^2 / (N_{osbl}^4 \, h^3).
\end{equation}
If the empirical rule of convection (equation (\ref{eq:empiricalRuleConvection})) is enforced in~(\ref{interm_eqn}), the result can be solved for $V_t^2$ so that 
\begin{equation}
V_t^2 = \frac{ \sqrt{0.2} \,  C_v}
    {\mbox{Ri}_{\mbox{\tiny crit}} \, \kappa^{2/3}} 
 \left(c_s \, \epsilon \right)^{-1/6} \, h \, N_{osbl} \, w_*
 \label{eq:vt2}
\end{equation}
where LMD94 included the constant $C_v$ to account for smoothing of the buoyancy profile near $h_e$ due to mixing.  

Note here that $N_{osbl}$ is the stratification near $h_e < h$. This approach contrasts to the use of stratification below the OSBL as suggested by \cite{Danabasoglu2006} (their Appendix A).  Both options are tested relative to LES in Sections~\ref{section:modifyNosbl}.  

\subsubsection{Enhanced diffusivity}
\label{enhancedDiffusivityDescribe}

For coarse vertical resolution, LMD94 propose the addition of an enhanced diffusivity to partially mitigate resolution-dependent biases. Here we suggest that the enhanced diffusivity parameterization also serves to parameterize unresolved sources of entrainment.

In their Appendix D, LMD94 note that as vertical resolution coarsens near the OSBL base, "staircase" structures can emerge in the time series of the boundary layer depth. We show this behavior in  Figure~\ref{matchconfig_compare}a, which is discussed in Section \ref{section:RTresults}. To understand the cause of the staircase structures, start by assuming the OSBL base is aligned with a grid cell base.  Furthermore, as in the case discussed in LMD94 (and the free convection test case described in Section \ref{subsection:test-case-introduction}), assume zero vertical property gradients within the boundary layer. Now allow the OSBL to deepen further into a new grid cell. As it deepens, boundary layer induced mixing within the new grid cell is not possible until the boundary layer reaches the bottom of this new cell. Once the OSBL depth reaches down to the next model interface, diffusivities become large, in which case properties mix quickly.  The resulting time series of boundary layer properties thus exhibits a stair-step structure. 

Use of a quadratic interpolation\footnote{In the CVMix and \cite{Danabasoglu2006} implementations of KPP, the stencil for quadratic interpolation includes the first model level where $Ri_b > Ri_{crit}$, as well as the two model levels above that layer.} to determine the OSBL depth has reduced these staircase structures \citep{Danabasoglu2006}.  Nonetheless, we still see such structures in our simulations (\textit{e.g.}, Figure~\ref{matchconfig_compare}a). 
Fundamentally, the staircase structures in OSBL depth illustrate a lack of an appropriate representation of unresolved entrainment fluxes.  The $V_t^2$ parameterization attempts to represent boundary layer entrainment, but given the dependence of $V_t^2$ on the modeled stratification near the OSBL base~(see Equation~(\ref{eq:vt2})), entrainment in KPP will be strongly resolution dependent. The LMD94 enhanced diffusivity parameterization also acts to enhance entrainment fluxes (Section~\ref{section:PEresults}) by increasing the KPP predicted diffusivity near the OSBL base, thus smoothing the boundary layer's vertical movement at coarser resolutions between grid cells. See Appendix D of LMD94 for further details regarding the specifics of the KPP enhanced diffusivity parameterization.

\subsection{The KPP parameterized non-local tracer transport}
\label{subsection:non-local-transport}

The parameterized non-local tracer flux from KPP is the product of the diffusivity (equation~(\ref{KPPDiffusivityText})) and a parameterized non-local term. The non-local term is non-zero only for tracers (though see \citet{Smyth2002} who suggest a form for momentum) in unstable surface buoyancy forcing, and it takes the form 
\cite{Mailhot1982}, LMD94)
\begin{equation}
\label{eq:gammab-defined}
\gamma_\psi = C_* \, \kappa \, \left(c_s \, \kappa \, \epsilon \right)^{1/3} 
 \left( 
 \frac{\overline{w^\prime \psi^\prime}_{\mbox{\tiny sfc}}}{w_\psi \, h} 
 \right), 
\end{equation}
where $\kappa$ is the von K\'arm\'an constant and $c_s$ and $C_{*}$ are constants (see LMD94 or \cite{Griffies2015} for details of these constants).  The turbulent tracer flux $(\overline{w^\prime \psi^\prime}_{\mbox{\tiny sfc}})$ arises from surface tracer transport due to air-sea or ice-sea interactions. Thus the non-local potential temperature flux ($\psi = \theta$), which is the product of the diffusivity and the non-local term, is given by
\begin{equation}
\label{non-localFlux}
K_\theta \, \gamma_\theta \equiv 
 \overline{w^\prime \theta^\prime}_{\mbox{\tiny non-local}} 
 = C_* \, \kappa \, \left(c_s \, \kappa \, \epsilon\right)^{1/3} G(\sigma) \, \overline{w^\prime \theta^\prime}_{\mbox{\tiny sfc}}.
\end{equation}
Hence, the parameterized non-local tracer flux is proportional to the shape function times the surface turbulent tracer flux. Specifically, this form for the KPP non-local potential temperature flux identifies it as a vertical redistribution throughout the boundary layer of the surface boundary flux $(\overline{w^\prime \theta^\prime}_{\mbox{\tiny sfc}})$.  The KPP non-local fluxes for other tracers take on the same form.  This form provides a useful conceptual framework for understanding the KPP non-local parameterization as well as a guide towards its numerical implementation.  

\subsubsection{Shortwave radiation and non-local heat transport}

The surface potential temperature flux includes penetrating shortwave radiation and longwave radiation. However, it is unclear how much of the shortwave absorbed in the boundary layer to include in $\overline{w^\prime \theta^\prime}_{\mbox{\tiny sfc}}$. The choice is  important as it impacts the strength of the non-local term. For example, use of all of the incident shortwave in $(\overline{w^\prime \theta^\prime})_{\mbox{\tiny sfc}}$ results in smaller magnitude for the destabilizing buoyancy forcing and thus a smaller non-local term.  We examine this choice in section~\ref{section:NLresults}.

\subsubsection{The non-local term and turbulence closure theory}

The derivation of the non-local tracer transport parameterization in equation~(\ref{eq:gammab-defined}) follows from a consideration of the turbulent scalar flux budget equation, which for temperature is given as
\begin{equation}
\label{temperatureFlux}
\frac{\partial \overline{w^\prime \theta^\prime}}{\partial t} =  -\underbrace{\overline{w^{\prime^2}} \frac{\partial \overline{\theta}}{\partial z}}_\textrm{local} + \underbrace{g \left(\alpha_\theta \overline{\theta^{\prime^2}} - \beta_S \overline{\theta^\prime S^{\prime}}\right)}_\textrm{buoyancy} - \underbrace{\frac{\partial \overline{w^\prime w^\prime \theta^\prime}}{\partial z}}_\textrm{triple moment} - \underbrace{\overline{\frac{\theta^\prime}{\rho_o}\frac{\partial p^\prime}{\partial z}}}_{\substack{\textrm{pressure-temperature}\\
\textrm{covariance}}}.
\end{equation}
The KPP non-local transport term follows the atmospheric modeling literature \citep{ertel1942vertical,Deardorff1966,Deardorff:1972,Mailhot1982,Holtslag1991} where in the budget for the turbulent vertical flux of potential temperature, the buoyant production term only involves the potential temperature variance.  Given these atmospheric derivations, the non-local tracer transport in KPP is truly only applicable to the vertical turbulent flux of buoyancy.  To parameterize a non-local transport of temperature and salinity, KPP assumes that the buoyancy flux separates cleanly into a potential temperature and salinity flux components via a linear equation of state and that the cross correlation of potential temperature and salinity is negligible in the buoyancy production term of equation~(\ref{temperatureFlux}). With these two assumptions, KPP can compute a non-local transport when there is a destabilizing surface flux of either heat or salt. We examine the KPP non-local transport term relative to LES in Section~\ref{section:NLresults}.

\subsubsection{Parameterized non-local salt transport}

LMD94 formulated KPP for models with a virtual salt flux at the ocean surface. For a virtual salt flux, the non-local salt fluxes are directly implemented just as the non-local heat fluxes given by equation~(\ref{non-localFlux}). However, in models making use of a freshwater surface boundary condition rather than a virtual salt flux\footnote{In this paper, MPAS-O and MOM6 use natural boundary conditions where fresh water is exchanged at the ocean-atmosphere interface, whereas POP uses a virtual salt flux.}, then salinity in the top layer is changed via surface mass fluxes ($Q_m$ arising from precipitation, evaporation, runoff, ice melt/formation).  Implementation of the non-local transport for real water flux models requires a separately defined salt flux given by  
\begin{equation}
\label{eq:Qs}
 Q_s = Q_m \, S_{\mbox{\tiny sfc}},
 \end{equation}
 where $S_{\mbox{\tiny sfc}}$ is the salinity in the surface grid cell. Equation~(\ref{eq:Qs}) also applies to all tracers that, like salt, have a high solubility in water and that do not leave the ocean with evaporating water.  

\subsection{Choices made by the circulation model}
\label{section:modelchoices}

We implemented KPP in three distinct ocean circulation models. In comparing results, we discovered that certain choices made by the circulation model can influence the results of the one-dimensional KPP tests. For example, mixing at the boundary layer base results in the transport of properties across the boundary layer base.  Therefore, it is expected that mixing and vertical grid resolution near the OSBL base will greatly influence the boundary layer properties. 

Furthermore, as discussed in the Section~\ref{section:entrainment}, the KPP $V_t^2$ parameterization is constructed to improve the representation of the entrainment buoyancy flux.  However, given the dependence on stratification near the OSBL, we expect the effectiveness of this parameterization to be sensitive to the chosen model resolution.  Therefore, we examine sensitivity to model resolution across MPAS-O, MOM6, and POP.  These tests include the use of constant vertical grid spacing and a non-uniform or stretched grid.  Our comparisons to LES confirm that there is indeed a resolution dependent bias.  We discuss this bias and possible remedies in Section~\ref{section:vertical-resolution}.

The three circulation models compute the vertical mixing tendency implicitly in time to avoid numerical stability restrictions on the timestep.  However, there can be biases that grow as the timestep changes due to a sensitivity to the parabolic Courant number \citep{Lemarie_etal2012a}.  Indeed, \cite{Reffray:2015kn} has found that certain turbulent closure schemes exhibit a fairly strong sensitivity timesteps.  Conversely, our results suggest that the KPP solution is robust to the chosen timestep (Section~\ref{section:timestep}).

\section{The LES model and the one-dimensional test cases}
\label{test_cases}

In this section, we present the LES model used for establishing a baseline behavior to be compared with KPP.  We then summarize the one-dimensional oceanic test cases used to benchmark KPP.

\subsection{Large Eddy Simulation (LES) model}

We compare one-dimensional column tests of KPP to a mature and commonly utilized LES model \citep{Moeng1984,McWilliams1997,sullivan2007surface}.  We have made two key modifications to the LES model for use in our tests.  First, we include salinity by setting the buoyancy equal to 
\begin{equation}
b = -g \left[ 1 - \alpha_{\theta} \left(\theta - \overline{\theta} \right) + \beta_S \left(S - \overline{S} \right) \right].
\label{eq:buoyancy}
\end{equation}
In this equation, $\alpha_{\theta} > 0$ is the thermal expansion coefficient, $\beta_S > 0$ is the haline contraction coefficient, and the overline represents a horizontal average over the domain. For our tests, we choose the constant values 
\begin{subequations}
\begin{align}
 \alpha_{\theta} &= 2 \, \times \, 10^{-4}~\mbox{K}^{-1}
 \label{eq:alphaT}
 \\
 \overline{\theta} &= 298.15 \mbox{K} \label{thetaBar}\\
 \beta_S &= 8\, \times \, 10^{-4}~\mbox{ppt}^{-1} \label{eq:betaT}\\
 \overline{S} &= 35 \mbox{ppt} .
\label{eq:sbar}
\end{align}
 \end{subequations}
In addition, the buoyancy production terms in the sub-grid TKE scheme have been modified so that 
\begin{equation}
\overline{w^\prime b^\prime} = g \left( \alpha_{\theta} \, \overline{w^\prime \theta^\prime} - \beta_S \, \overline{w^\prime S^\prime} \right).
\label{eq:wb}
\end{equation}
Next, the stability factor in the length scale parameterization \citep{deardorff1980stratocumulus} has been modified, with evaluation of these changes presented in Appendix~\ref{salinity:testing}.

An additional modification involves the implementation of a diurnal cycle in the LES model along with shortwave radiation (similar to \citealp{Wang1998}).  The vertical penetration of incoming solar radiation follows a two-band exponential formulation with constant extinction coefficients of a Jerlov type IB water mass \citep{Paulson1977}, which is consistent with that used in MPAS, POP, and MOM6.  The function describing the time variation of surface shortwave radiation is described in Appendix~\ref{shortwave:desc}.

In most simulations, the LES utilizes a stretched grid over its 150~m vertical extent, with the first layer thickness of 0.1~m.  The LES has a horizontal extent of 128 m and a uniform horizontal resolution of 0.5 m.  For the surface heating and wind forced case, the LES resolution is uniform in the horizontal and vertical at 0.25m.

\subsection{Description of the test cases}
\label{subsection:test-case-introduction}

We examine the impacts of the implementation considerations discussed in Sections~\ref{sec:shapefunc-diffusmatch} -~\ref{section:modelchoices} through a series of one-dimensional test cases.  The tests span a range of oceanographically relevant forcing and include
\begin{itemize}
\item Convective deepening induced by surface cooling with no initial OSBL (linearly stratified in temperature),
\item OSBL deepening dominated by wind stress with no initial OSBL,
\item OSBL deepening via wind stress with surface heating
\item OSBL deepening generated by mechanical and thermodynamic forcing with no initial OSBL,
\item Turbulence generated by surface cooling with a preexisting thermocline and halocline,
\item The influence of boundary layer deepening into preexisting background shear, and
\item The influence of diurnal variability in surface buoyancy forcing.
\end{itemize} 

To easily identify each test in discussion, the test are named according to the salient details.  For example, the test forced by surface cooling, evaporation, and wind is designated CEW.  Table~\ref{testNames} show the names and essential details for each simulation.  Full details of each simulation configuration and forcing are given in Tables~\ref{exp_params} and \ref{TS_profiles}.

\begin{table}
	\centering
	\def\arraystretch{1.5}
	\begin{tabular}{ l  c }
		\hline
		Test Name & Salient Details \\
		\hline
		FC & Free Convection \\
		CEW & surface Cooling, Evaporation, and Wind stress  \\
		FCML & Free Convection with T \& S Mixed Layers  \\
        WNF & Wind stress with No Coriolis   \\
        CWB & surface Cooling and Wind stress with Background shear \\
        FCE & Free Convection due to surface Evaporation \\
        DC & Diurnal Cycle  \\
        HW & surface Heating and Wind stress  \\
        \hline
	\end{tabular}
	\caption{Test case names and salient features.} 
	\label{testNames}
\end{table}

Note that our focus is on the OSBL portion of KPP and not the specifics of interior mixing.  Certainly the interior mixing scheme will influence the OSBL via diffusivity matching in KPP.  In cases with surface momentum forcing and/or background shear, the LMD94 shear instability scheme will be used.  This parameterization computes diffusivities as a function of the gradient Richardson number.

In most test cases, the models are initialized with zero velocity. However, in CWB, a constant background shear layer is imposed via a background velocity profile given as:
\begin{equation}
\label{eq:background_shear}
 \overline{U}(z) =\begin{cases}
    0, & \text{if $z>-25$},\\
    -\frac{0.3}{50.0}(z+25), & \text{if $z\leq-25$ and $z > -75$}, \\
    0, & \text{otherwise}.
  \end{cases}
\end{equation}
Here, the imposed shear is roughly equivalent to observed shear in the equatorial undercurrent \citep{johnson2002direct}.

In all simulations, we define a base configuration of KPP consistent with \cite{Danabasoglu2006}.  The base configuration of KPP is summarized in Table~\ref{defaultKPPconfig}.  Note that in our base configuration, the matching of internal diffusivity gradients is abandoned.  Sensitivity to this choice is examined in Section \ref{section:DMresults}. 

In the FC, FCML, and DC tests all internal mixing schemes (mixing below the boundary layer) are disabled, so the diffusivity vanishes beneath the boundary layer. Hence, any diffusivity  matching (Section \ref{sec:shapefunc-diffusmatch}) from LMD94 has no influence.   

A number of specifications common to most test cases are as follows:

\begin{sidewaystable}
	\def\arraystretch{1.5}
	\centering
	\begin{tabular}{ l  c  c  c  c  c  c  c  c  c }
		\hline
		Simulation & $ Q_o (W m^{-2}) $ & $Q_{sw}^{max} (W m^{-2}) $ & E (mm/day)& $\tau_x (Pa) $ &  $T(z), S(z)$ & $f (s^{-1})$ \\
		\hline
		FC & $ -75 $ &  0 & 0 & 0  & A & $10^{-4}$\\
		CEW & $-75$ &  0 & $1.37$ & 0.1 & A & $10^{-4}$\\
        FCML & $-75$ &  0 & 0 & 0 & C & $10^{-4}$\\
		WNF & $ 0 $ &  0 & $0$ & 0.1 & D & 0\\
		CWB & $-75$ &  0 & $0$ & 0.1 & E & 0 \\
		FCE & 0 &  0 & $1.37$ & 0 & B & $10^{-4}$ \\
		DC & $-75$ & $235.62$ & 0 & 0 & A & $10^{-4}$\\
        HW & $75$ & 0 & 0 & 0.1 & A & $10^{-4}$ \\
		\hline
	\end{tabular}
	\caption{Summary of forcing scenarios considered in the test cases. $Q_o$ is the non-solar surface heat flux (positive is into the ocean), and $Q_{sw}^{max}$ is the maximum of surface diurnal shortwave radiation. We provide details of the diurnal shortwave forcing in Appendix \ref{shortwave:desc}. E is the surface evaporation rate, $\tau_x$ is the zonal wind stress, and $\tau_y = 0$ in all simulations. $T(z)$ and $S(z)$ are the initial temperature and salinity profiles, which are given in Table~\ref{TS_profiles}. The Coriolis parameter ($f$) is held fixed for a given simulation. In CWB, a background shear is imposed (equation~(\ref{eq:background_shear})) The FCE test is used to verify the LES implementation of salinity in Appendix~\ref{salinity:testing}.} 
\label{exp_params}
\end{sidewaystable}

\begin{table}
	\centering
	\def\arraystretch{1.5}
	\begin{tabular}{ l  c  c  }
		\hline
		Profile & $T(z)$ & $S(z)$ \\
		\hline
		A & $20 + 0.01z$ & 35  \\
		B & 20 & $35 - 0.007813z$  \\
		C & $\begin{array} {lll} 20 \qquad & \mbox{if} \; z \geq -25 \\  20 + 0.01 (z + 25) \qquad & \mbox{else} \end{array}$ & $\begin{array} {lll} 35 \qquad & \mbox{if} \; z \geq -25 \\ 35 - 0.03(z+25) \qquad &\mbox{if} \; -25 \geq z \geq -35 \\ 35.3 \qquad &\mbox{else} \end{array}$  \\
        D & $20 + 0.05z$ & 35  \\
        E & $\begin{array} {lll} 20 \qquad & \mbox{if} \; z \geq -25 \\  20 + 0.05 (z + 25) \qquad & \mbox{else} \end{array}$ & $\begin{array} {lll} 35 \end{array}$  \\
		\hline
	\end{tabular}
	\caption{Initial temperature ($\mbox{}^{\circ}\mbox{C}$) and salinity (ppt) profiles used in the experiments given in Table~\ref{exp_params}.  Recall that the vertical position (geopotential coordinate $z$) is  positive upwards, with $z=0$ at the resting ocean surface and $z<0$ in the ocean interior.} 
	\label{TS_profiles}
\end{table}

\begin{itemize}
\item MPAS, MOM6, and POP all apply diffusive tendencies implicitly, while the LES uses an explicit discretization.
\item We use a linear equation of state (equation~(\ref{eq:buoyancy})). The values of $\alpha_{\theta}$ and $\beta_S$ used in equation~(\ref{eq:buoyancy}) are identical to those used for the LES (see equations~(\ref{eq:alphaT}) and~(\ref{eq:betaT})).  Given a linear equation of state, the surface buoyancy flux is given by  
\begin{equation}
\label{eq:wbSFC}
 \overline{w^\prime b^\prime}_{\mbox{\tiny sfc}} = 
 g 
 \left(
 \alpha_{\theta} \, 
 \overline{w^\prime \theta^\prime}_{\mbox{\tiny sfc}} 
 -\beta_S \, 
 \overline{w^\prime S^\prime}_{\mbox{\tiny sfc}}
 \right).
 \end{equation}
 Hence, a positive correlation between vertical velocity fluctuations and temperature fluctuations contribute to a positive turbulent buoyancy flux, as does a negative correlation between vertical velocity and salinity fluctuations. 

\item A variety of vertical grid spacings are considered for the KPP tests: some with uniform vertical spacing (1~m to 10~m) and one with non-uniform grid spacing (1.5 m near surface).  

\item Most simulations are run for eight days.  WNF is run for only one day, the CWB is run for four days, and the FCML test is run 12 days.

\item In the DC test, the time dependent shortwave heat flux is constructed such that the daily integrated positive (stabilizing) buoyancy input is balanced by the daily integrated negative (destabilizing) buoyancy flux.  The explicit form of the shortwave radiation used in the DC test given in equation~(\ref{eq:qswMax}) (see Appendix~\ref{shortwave:desc}) and the maximum daily shortwave radiation is listed in Table~\ref{exp_params}. 

\item Across all configurations and parameter settings more than 100 tests were conducted.

\end{itemize}
Finally, we note that most of the test cases outlined here expose more than one issue discussed in Section~\ref{issues}.  Table~\ref{TestCases} lists the four focus areas illustrated in  Figure~\ref{BL_schematics} as well as the tests that are pertinent to each implementation consideration.  Within many test cases, numerous sensitivities are examined. In Table~\ref{parameter_tests}, we summarize the labels used for each sensitivity test and changes made to KPP relative to the baseline configuration of Table~\ref{defaultKPPconfig}.

\begin{table}
	\centering
	\def\arraystretch{1.5}
	\begin{tabular}{ l  c  c }
		\hline
		 & \begin{tabular}{@{}c@{}}KPP description \\ and results sections \end{tabular}  &  Pertinent Tests \\
		\hline
		Diffusivity Matching & \begin{tabular}{@{}c@{}}Section~\ref{sec:shapefunc-diffusmatch}\\ Section~\ref{section:DMresults}\end{tabular} & CEW, FCML  \\
		Property Exchange & \begin{tabular}{@{}c@{}}Section~\ref{section:entrainment} \\ Section~\ref{section:PEresults} \end{tabular}& \begin{tabular}{@{}c@{}}FC, CEW, FCML, \\ CWNF, DC \end{tabular}  \\
		Non-local Transport & \begin{tabular}{@{}c@{}} Section~\ref{subsection:non-local-transport} \\ Section~\ref{section:NLresults} \end{tabular} & CEW, FCML, CWB  \\
        Resolution and Timestep & \begin{tabular}{@{}c@{}} Section~\ref{section:modelchoices} \\ Section~\ref{section:RTresults} \end{tabular} & FC, CEW, WNF  \\
        \hline
	\end{tabular}
	\caption{List of Theory and Results section pertinent to each implementation consideration shown in Figure~\ref{BL_schematics}. The final column lists each test case pertinent to each individual issue.} 
	\label{TestCases}
\end{table}

\begin{table}
	\def\arraystretch{1.5}
	\centering
	\resizebox{\columnwidth}{!}{

	\begin{tabular}{ l  c  }
		\hline
		 Test label & Parameter(s) changed  \\
		\hline
		Base & follows Table~\ref{defaultKPPconfig} \\
        NM & internal matching disabled \\
        N2 & $N = \max(N(k_{osbl-1}),N(k_{osbl}))$ as in equation~(\ref{eq:newN-defn})  \\
        MB & match $\partial_z K_\psi^{int}(h)$ in internal matching\\
        \hline
	\end{tabular}
	}
	\caption{Sensitivity tests conducted within many of the test cases detailed in Tables~\ref{exp_params} -~\ref{TS_profiles}.  Test labels correspond to figures in Sections~\ref{section:RTresults}-\ref{section:NLresults}} 
    \label{parameter_tests}
\end{table}

\begin{table}
	
	\def\arraystretch{1.5}
	\centering
	\resizebox{\columnwidth}{!}{
	\begin{tabular}{ l  c  }
		\hline
		 {\sc parameterization choice} & {\sc default value}  \\
        \hline
		OSBL interpolation & quadratic  \\
		Timestep & 20 min \\
        Internal matching  & Match Diffusivity only  \\
        Interpolation order for internal matching & linear \\
        Convective diffusion & enabled  \\
        $\mbox{Ri}_{\mbox{\tiny crit}}$ 
        (critical Richardson number) & 0.25 \\
        $\epsilon$ (fraction of OBSL occupied  by surface layer) & 0.1 \\
		Enhanced diffusivity at OSBL base & enabled \\
        Shear instability mixing & LMD94 scheme \\
        Background diffusivity & zero \\
        Internal wave mixing &	zero \\
        Double diffusion & zero  \\
        \hline
	\end{tabular}
	}
	\caption{Summary of the baseline KPP configuration. Note that in all runs, a large vertical diffusivity meant to parameterize convection is computed after KPP and only below the OSBL.} 
    \label{defaultKPPconfig}
\end{table}

\section{Sensitivities to the space and time discretization} 
\label{section:RTresults}

In this section, we examine the performance of the CVMix version of KPP as realized in three ocean circulation models.  We thus examine the sensitivity of KPP results to certain model details such as horizontal grid layout, vertical grid resolution, and time step size.  Our tests generally show that KPP is quite sensitive to vertical resolution but is relatively insensitive to horizontal discretization and time step size.

\subsection{Sensitivity to horizontal discretization}

We here incorporate KPP into three ocean circulation models.  Notable differences in the models relate to their choice for horizontal discretization. POP utilizes a staggered B-grid, while MPAS-O and MOM6 utilize the C-grid \citep{arakawa1981potential}.  Furthermore, POP and MOM6 utilize a structured quadrilateral grid, whereas MPAS utilizes unstructured Voronoi meshes \citep{ringler2013multi}. In principle, the horizontal discretization should not matter since KPP is a one-dimensional scheme acting in each vertical column. However, there are quantities needed by KPP that require horizontal information for averaging onto the tracer cell (e.g., vertical shear of the horizontal currents needed by the bulk Richardson number in equation (\ref{eq:rib-defined})). We thus do not presume, before testing, that results are robust across the models.   

\begin{figure}
\centering\includegraphics[width=0.9\linewidth]{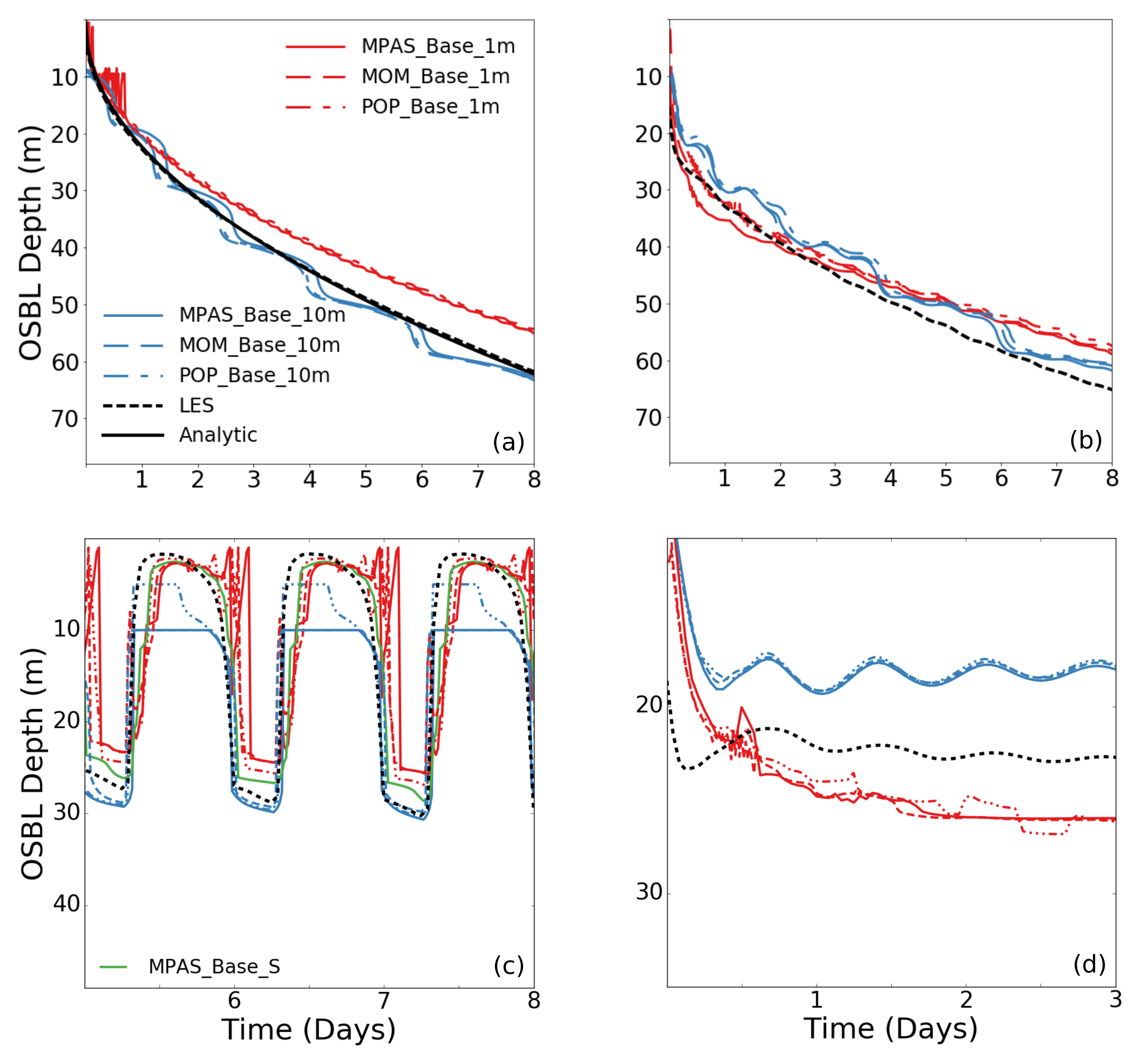}
\caption{Boundary layer depths computed using equation~(\ref{eq:KPP-boundary-layer-depth-defined}) from the base configuration of MPAS-O, MOM, and POP. (a) Free Convection (FC) test, (b) Convection, Evaporation, and Wind (CEW) test, (c) Diurnal Cycle (DC) test, and (d) Heating and Wind (HW) test.  In each panel, the LES output is dashed black and in (a) the thick black line is an analytic solution computed from equation~(\ref{finalH}).}
\label{matchconfig_compare}
\end{figure}

In Figure~\ref{matchconfig_compare}, we show the KPP OSBL depths computed using the baseline configuration for the free convection (FC), CEW (cooling, evaporation, wind), diurnal cycle (DC), and HW (heating, wind) tests for MPAS-O, MOM6, and POP, as well as for the LES and an analytic solution (see equation~(\ref{finalH}) in Appendix~\ref{analytic}) in the FC test. The three circulation models yield very similar OSBL depths, yet they systematically deviate from the LES solution in the FC and HW tests.  The largest difference between the models is in the DC test for shallow boundary layers in stable forcing, which is due to the assumed minimum boundary layer depth.  In POP the minimum OSBL depth is half of the first model thickness whereas MPAS and MOM6 choose the full layer thickness.  

\begin{figure}[thbp]
\centering\includegraphics[width=0.6\linewidth]{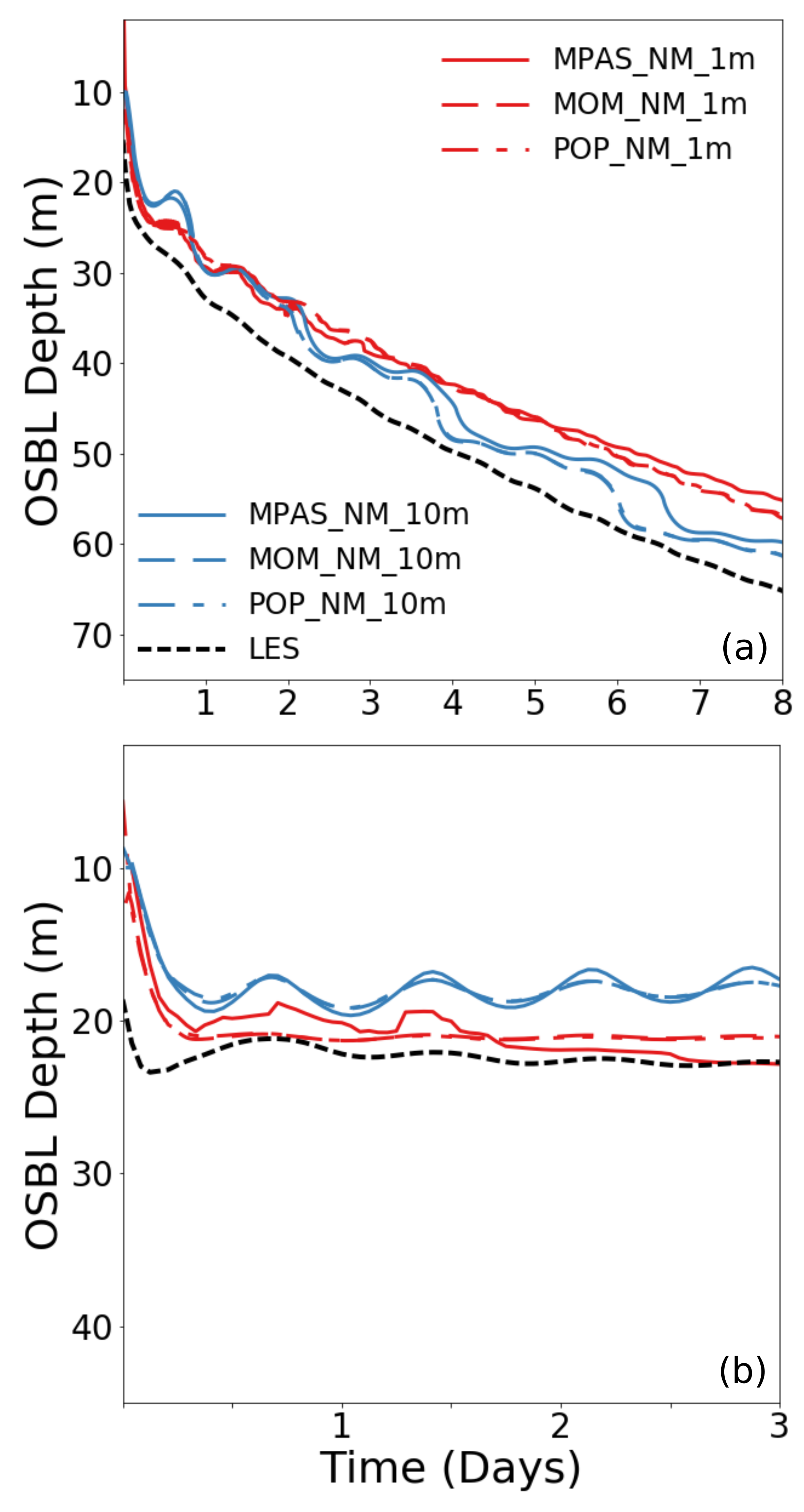}
\caption{As in Figure~\ref{matchconfig_compare} but for the no internal match configuration.  (a) CEW test and (b) HW test.  The FC and DC tests are not shown as there is no predicted internal mixing and thus the solutions are identical for the baseline and no match configurations.}
\label{nomatch_coparison}
\end{figure}

We also have tested the no match configuration (where internal matching is abandoned) in the three circulation models for two of the four tests.  In the FC and DC tests the contribution from interior mixing is zero and hence the no match and baseline configurations are identical.  The predicted OSBL depths for the HW and CEW tests are shown in Figure~\ref{nomatch_coparison}.  As in the baseline configuration the results from the three models are consistent.  In the HW test, the high resolution results have a smaller OSBL bias relative LES than coarse resolution (compare Figure~\ref{matchconfig_compare}d).  The improvement in the no match configuration is discussed in Section~\ref{section:osbl-base-treatment}.

Given that MPAS-O, MOM6, and POP are consistent across tests that span important forcing regimes (convective, shear driven, and stable heating) and KPP configurations, we utilize MPAS-O for various sensitivity tests in subsequent sections.

\subsection{Exhibiting sensitivity to vertical grid resolution}
\label{section:vertical-resolution}

A number of the test cases (e.g. FC and DC) show that, surprisingly, the KPP boundary layer depth is more consistent with LES at coarser resolution.  In contrast, the finer grid resolution simulations exhibit a persistent shallow boundary layer bias. As an example, Figures~\ref{matchconfig_compare}a and c illustrate a persistent shallow bias in OSBL depths at high resolution relative to LES and coarse resolution KPP results.  

To further quantify the resolution dependent bias, we have examined two additional vertical grids: d$z = 20$m in the FC and FCML (free convection, mixed layer) tests and a non-uniform grid with 1.5~m resolution near the surface (tagged S in Figures~\ref{matchconfig_compare} and~\ref{n2_osbl_comparison}) in the diurnal cycle (DC) test.  Such non-uniform grid spacing is commonly used in realistic climate models.  The resulting OSBL depths are shown in Figure~\ref{n2_osbl_comparison}.  In the FC and FCML tests and during times of destabilizing surface buoyancy in the DC test there is a persistent shallow bias at the finest grid spacing.  As the grid spacing coarsens, the OSBL deepens.  When d$z = 20$m the OSBL is deeper than LES (Figure~\ref{n2_osbl_comparison}a and e), suggesting that the minimal bias seen for d$z = 10$m in the FC and FCML tests is fortuitous.

At the two finer resolutions (d$z = 1$m and stretched grid), high frequency temporal OSBL noise develops near the surface boundary during stabilizing buoyancy forcing (e.g., Figure~\ref{n2_osbl_comparison}b) in the DC test case.  This behavior is similar to what is seen in the FC test (Figure~\ref{n2_osbl_comparison}a, near the start of the simulation).  The noise at high, near surface resolution, and the deep bias for coarse resolution suggests that the enhanced diffusivity parameterization is an incomplete representation of unresolved entrainment across resolutions and surface forcing.  We examine these biases in more detail and present a possible solution in Section~\ref{section:PEresults}.

\subsection{Sensitivity to time step}
\label{section:timestep}

The WNF (wind stress with no Coriolis) test has been used as a benchmark for one-dimensional turbulence models (\textit{e.g.}, \citealp{burchard2001comparative}).  It is motivated by the laboratory experiment conducted by \cite{kato1969penetration}, who measured the deepening of the surface boundary layer in an initial linearly stratified fluid ($N^2=10^{-4} s^{-2}$) forced by a constant surface friction velocity ($u_* = 0.01 ms^{-1}$).  

The boundary layer depth in the experiment is defined as the depth of the maximum $N^2$ in the water column.  We refer to this as the Kato Phillips (KP) boundary layer depth ($h_{\mbox{\tiny KP}}$). \cite{kato1969penetration} found that the KP boundary layer depth followed an empirical relation given by
\begin{equation}
h_{\mbox{\tiny KP}} = \frac{1.05\, u_* \, \sqrt[]{t}}{\sqrt[]{N(t=0)}}.
\label{eq:KPmld}
\end{equation}
The WNF test is run for 24 hours as \citet{kato1969penetration} found this relation to only be valid for timescales on the order of 30 hours.

\begin{figure}[hbtp]
\centering\includegraphics[width=0.99\linewidth]{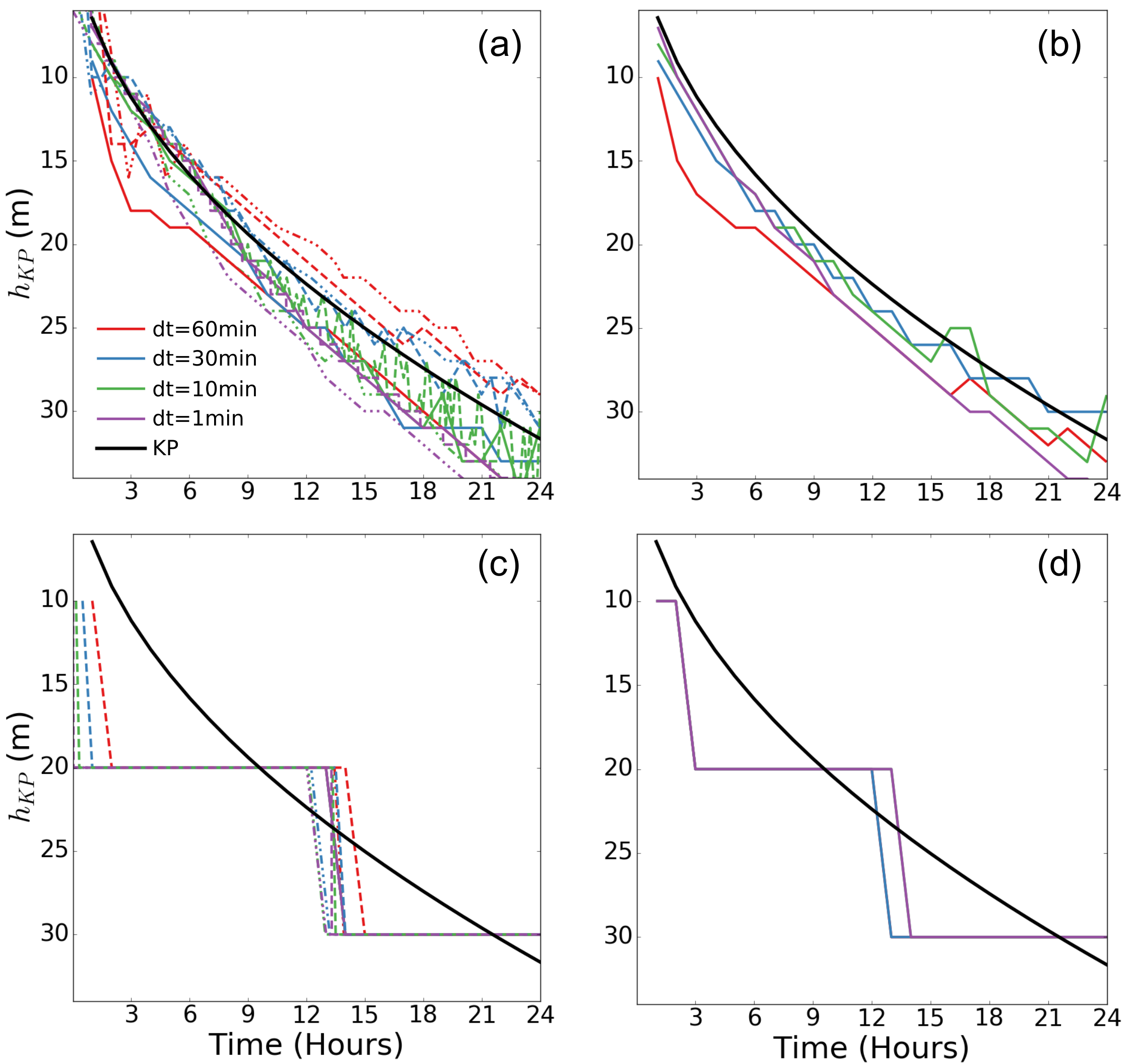}
\caption{Test Case WNF: Evolution of $h_{\mbox{\tiny KP}}$, defined in equation~(\ref{eq:KPmld}).  (a) Base configuration -- d$z = 1$m, (b) No match configuration with d$z = 1$m, (c) Base configuration -- d$z = 10$m, and (d) No match configuration -- d$z = 10$m.  In every plot, the colors represent the time step size in minutes (see legend in (a)).  The black line in all plots is given by equation~(\ref{eq:KPmld}).  In the base configuration, the solid lines are MPAS results, the dashed line are MOM6 results, and the dot-dashed lines are POP results.  See Section \ref{section:DMresults} for discussion of the no matching approach.}
\label{wind1_bld}
\end{figure}

Figure~\ref{wind1_bld} shows the simulated $h_{\mbox{\tiny KP}}$ for the base (a and c) and the no match case (b and d) configurations of KPP.  At fine resolution and across all timesteps tested, KPP (in both configurations) well captures the empirical relation given in equation~(\ref{eq:KPmld}). Coarse resolution simulations also perform reasonably well. Consistent with Section~\ref{section:vertical-resolution}, there are only minor differences between MPAS and MOM6 results.  We attribute the noise in Figure~\ref{wind1_bld}a to the algorithm we use to diagnose $h_{\mbox{\tiny KP}}$.  In particular, we did not interpolate between levels. Rather, the algorithm simply returns the level of maximum stratification in the column.  Our results show that KPP is robust with respect to the chosen time step.

\section{Treatment of the boundary layer base}
\label{section:osbl-base-treatment}

The KPP diagnosed OSBL depth is a critical parameter in the predicted vertical turbulent flux (e.g. equations~(\ref{KPPfluxText}) and~(\ref{KPPDiffusivityText})).  The diagnosed boundary layer depth is dependent on the near surface processes through the surface layer buoyancy and momentum (equation~(\ref{eq:KPP-boundary-layer-depth-defined})) and processes near the boundary layer base.  Here we examine KPP sensitivities to the parameterization of entrainment and diffusivity near the boundary layer base.

\subsection{Remedies for the shallow bias with fine grid spacing}

We revisit how the boundary layer depth is diagnosed as a means to help understand the shallow bias seen in the fine-resolution KPP simulations. For simplicity, focus on the FC and DC tests, in which the KPP boundary layer depth computed according to equation~(\ref{eq:KPP-boundary-layer-depth-defined}) simplifies to
\begin{equation}
\label{eq:convect1-Rib}
Ri_{\mbox{\tiny crit}} = \frac{h \, \left[b_{\tiny \mbox{sl}} - b(h) \right]}{C \, h \, N_{osbl} \, w_*}.
\end{equation}
To reach this expression, we made use of a simplified definition of the unresolved shear, $V_t^2$, in which all the constants in equation~(\ref{eq:vt2}) are subsumed into $C$, and we further assume zero velocity. Hence, for a fixed boundary layer buoyancy, $b_{\tiny \mbox{sl}}$, and surface buoyancy forcing, which fixes $w_*$, the key means to modify the Richardson number and hence deepen the boundary layer ($h$) in equation~(\ref{eq:convect1-Rib}) is via the stratification near the OSBL base as measured by $N_{osbl}$.

\subsubsection{Sensitivity of the Bulk Richardson number to $N_{osbl}$}

Increased stratification near the OSBL base increases the unresolved turbulent shear, $V_t^2$ (equation~(\ref{eq:vt2})). This increased $V_t^2$ in turn reduces the bulk Richardson number. The diagnosed boundary layer depth then increases to further enhance mixing. At fine vertical resolution, the stratification varies rapidly in the entrainment layer, which is a region of large stratification (e.g., the region near $d/h=1$ in Figure~\ref{Vt_schematics}).  These features of the boundary layer and the adjacent interior suggest that the KPP boundary layer depth can be very sensitive to the chosen level of stratification used for computing $V_t^2$.

\cite{Danabasoglu2006} define $N_{osbl}$ for the Richardson number calculation according to 
\begin{equation}
\label{d06N}
N_{osbl} = \frac{b(k_{osbl+1/2}) - b(k_{osbl-1/2})}{z(k_{osbl+1/2}) - z(k_{osbl-1/2})},
\end{equation}
where $k_{osbl}$ is the vertical index for the grid cell closest to the OSBL depth. Yet, as shown in Section~\ref{section:vt2}, the chosen $N_{osbl}$ should be defined one cell shallower in the column, i.e., 
\begin{equation}
\label{alternateN}
N_{osbl} = \frac{b(k_{osbl-1/2}) - b(k_{osbl-3/2})}{z(k_{osbl-1/2}) - z(k_{osbl-3/2})}.
\end{equation}

\begin{table}
	\centering
	\def\arraystretch{1.5}
	\begin{tabular}{ l  c  c  c  c  }
		\hline
		Test Name & $V_t^2(h)_{orig} \, (m^2 s^{-2}) $ & $V_t^2(h)_{new} \, (m^2 s^{-2})$ \\
		\hline
        FC\_1m & 0.0147 & 0.0225  \\
        FC\_10m &  0.0173 & 0.0170 \\
        CEW\_1m & 0.0129 & 0.0192 \\
        CEW\_10m & 0.0124 & 0.0139\\
        FCML\_1m & 0.0354 & 0.0223 \\
        FCML\_10m & 0.0094 & 0.0148 \\
        HW\_1m & 0.0112 & 0.0042 \\
        HW\_10m &  0.0055 & 0.0039\\
        DC\_1m &  $1.50 \times 10^{-3}$ & $1.88 \times 10^{-3}$  \\
        DC\_10m & $2.17 \times 10^{-3}$ & $1.50 \times 10^{-3}$ \\
        \hline
        
	\end{tabular}
	\caption{Influence of alteration of $N_{osbl}$ on $V_t^2$.  The subscript $orig$ defines $N_{osbl}$ via equation~(\ref{d06N}) and subscript $new$ uses equation~(\ref{alternateN}). The 1m and 10m tags specify the vertical resolution.  All values are averaged over the last day of the simulation.} 
	\label{Table:N2_influence}
\end{table}

Table~\ref{Table:N2_influence} shows the influence of changing the definition of $N_{osbl}$.  The middle column defines $N_{osbl}$ via equation~(\ref{d06N}) and the third column uses equation~(\ref{alternateN}).  Tests and resolutions where $V_{t_{new}}^2$ is larger than $V_{t_{orig}}^2$ will exhibit deeper boundary layers (assuming identical surface forcing).  As expected, the influence of $N_{osbl}$ is smaller for coarser resolution as the entrainment layer is not well resolved.  Note that in many of the tests, the diagnosed value of $V_t^2$ decreases when $N_{osbl}$ is defined by equation~(\ref{alternateN}), which would result in shallower boundary layers.

\subsubsection{Modifying $N_{osbl}$ to reduce the shallow bias}
\label{section:modifyNosbl}
The above considerations suggest equation~(\ref{alternateN}) is not optimal. It is also not robust since it can, particularly at coarse resolution, lead to $N_{osbl} \approx 0$ which leads to a non-robust calculation of the Richardson number.

\begin{figure}[thbp]
\centering\includegraphics[width=0.85\linewidth]{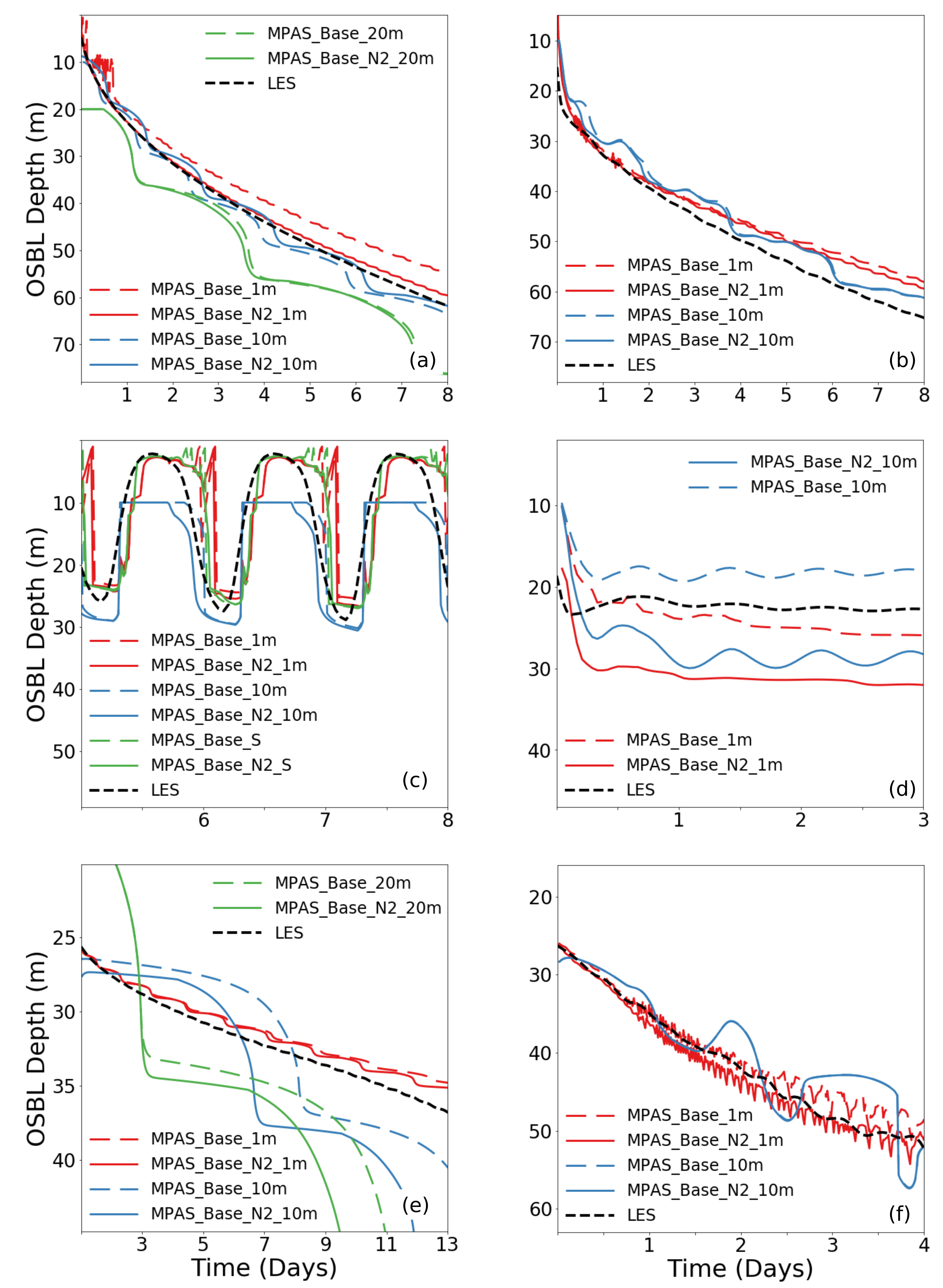}
\caption{KPP sensitivity to the chosen level of stratification in equation~(\ref{eq:vt2}).  In all panels, the base configuration result is reproduced as the dashed lines.  The solid lines are for tests using equation~(\ref{eq:newN-defn}) and are tagged N2.  (a) FC test, (b) CEW test, (c) DC test, (d) HW test, (e) FCML test, and (f) CWB test.  In all plots, the boundary layer depths are computed using equation~(\ref{eq:KPP-boundary-layer-depth-defined})}
\label{n2_osbl_comparison}
\end{figure} 

Our tests suggest the following more preferable alternative given by 
\begin{equation}
\label{eq:newN-defn}
N_{osbl} = \max\left(\frac{b(k_{osbl+1/2}) - b(k_{osbl-1/2})}{z(k_{osbl+1/2}) - z(k_{osbl-1/2})},\frac{b(k_{osbl-1/2}) - b(k_{osbl-3/2})}{z(k_{osbl-1/2}) - z(k_{osbl-3/2})}\right).
\end{equation}
In Figure~\ref{n2_osbl_comparison}, we exhibit OSBL depths from a series of tests using equation~(\ref{eq:newN-defn}) for the $V_t^2$ calculation.  The new definition of $N_{osbl}$ greatly reduces the shallow bias found at fine resolution in the FC test (Figure~\ref{n2_osbl_comparison}a). The coarse resolution (d$z = 10$m and d$z = 20$m) FC result is nearly unchanged from the base case, since the entrainment layer is strongly smoothed at coarse resolution and the stratification is constant below the boundary layer.

In the CEW and CWB tests, the model-predicted shear of horizontal momentum dominates the denominator of Equation~(\ref{eq:KPP-boundary-layer-depth-defined}), minimizing the influence of $V_t^2$ and hence $N_{osbl}$ (Figures~\ref{n2_osbl_comparison}b and f).

In the DC test, the new calculation of $N_{osbl}$ again slightly diminishes the resolution dependent bias during destabilizing surface buoyancy forcing (Figure~\ref{n2_osbl_comparison}c).  In the DC test we again present results from a non-uniform vertical resolution.  The influence of altering $N_{osbl}$ is weaker at this resolution and continues to weaken as the resolution coarsens.  Yet, unlike the FC test, the near-surface noise remains (compare Figures~\ref{n2_osbl_comparison}a and c). 

Under stabilizing surface forcing (the HW test), altering $N_{osbl}$ deepens the OSBL significantly.  The shallow OSBL bias at coarse resolution is now a deep bias.  The slight OSBL deep bias increases at high resolution.

In the FCML test (Figure~\ref{n2_osbl_comparison}e) the boundary layer is roughly unchanged when equation~(\ref{eq:newN-defn}) is used in the $V_t^2$ parameterization. Altering the definition of $N_{osbl}$ at coarse resolution  has a larger impact on the KPP simulated OSBL.  The dependence of KPP on $V_t^2$ at coarse resolution was not seen in most tests (Figure~\ref{n2_osbl_comparison}).  We believe that the coarse resolution KPP is more sensitive to the $V_t^2$ parameterization in the FCML test due to the very strong stratification near the base of the OSBL.

In most cases, altering the definition of $N_{osbl}$ reduces OSBL bias relative to LES, although slightly.  In the FCML and HW tests the OSBL bias increases.  Our results suggest that altering $N_{osbl}$ is a useful, but not critical consideration for KPP configurations.

\subsection{Diffusivity at the boundary layer base}
\label{section:DMresults}

As discussed in Section~\ref{sec:shapefunc-diffusmatch}, there are a number of important considerations regarding the internal matching scheme of KPP.  Our results suggest that the base and no match configurations, with a few critical alterations, can perform well relative to LES.

\subsubsection{Considerations for the no-match configuration}

\begin{figure}[thbp]
\centering\includegraphics[width=0.5\linewidth]{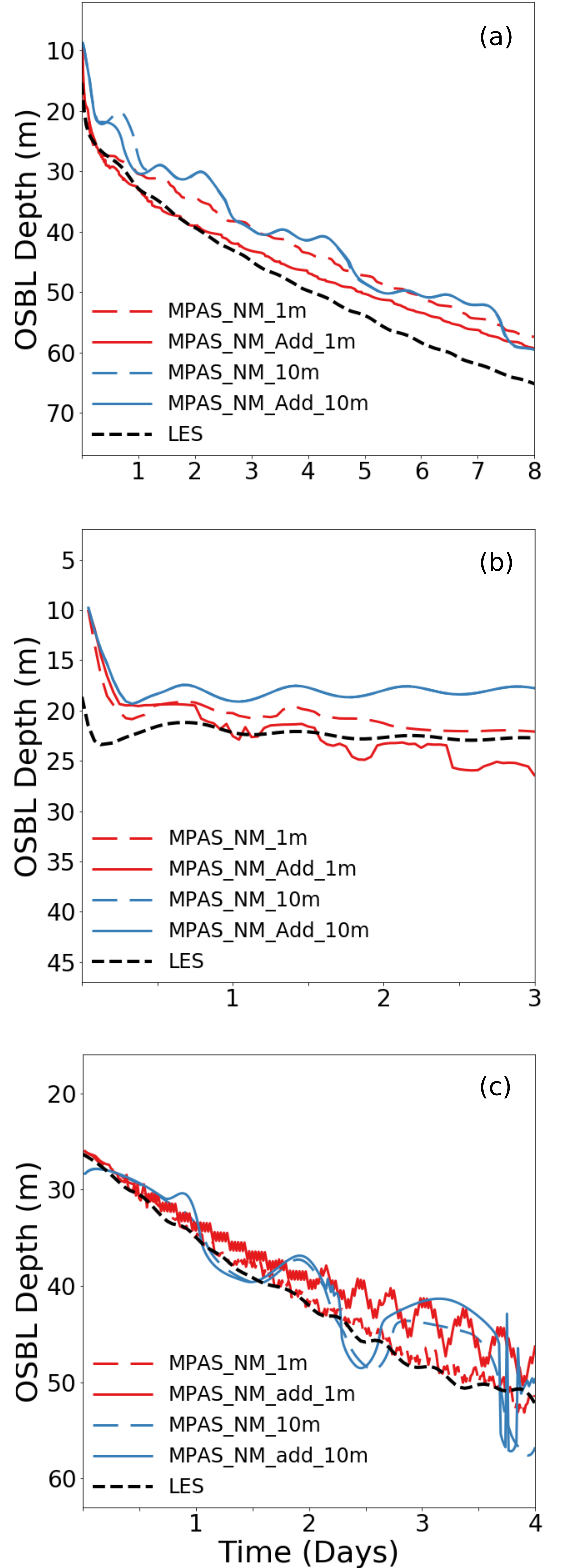}
\caption{OSBL depths from tests examining the sensitivity of adding shear instability mixing in the boundary layer when diffusivity matching is disabled. (a) CEW test, (b) HW test, and (c) CWB test.  In every test, the boundary layer depth is calculated following equation~(\ref{eq:KPP-boundary-layer-depth-defined}).
}
\label{fig:NM_v_add}
\end{figure}

In the convection, evaporation, and wind (CEW) and heating and wind (HW) tests, the OSBLs depth from the no match (NM) configuration are shallower than the base configuration of KPP (compare Figures~\ref{matchconfig_compare}b and d to Figure~\ref{nomatch_coparison}).  We suggest that the shallow OSBL bias in the NM configuration is partially caused by the inability of shear instability mixing to influence the OSBL (consistent with LMD94).  Figure~\ref{fig:NM_v_add} shows the influence of extending the LMD94 shear instability mixing scheme into the OSBL (NM\_Add test) for the three tests that are wind forced.  At high resolution, the OSBL depths increase when shear instability mixing is included in the OSBL, decreasing the bias relative to LES.  

In the convection, wind, background shear (CWB) test, which simulates a boundary layer deepening into a region of preexisting shear, the OSBL bias increases.  Temporal noise also develops.  

At coarse resolution, the inclusion of shear instability mixing in the OSBL does not dramatically alter the results.  It is likely that the insensitivity to interior mixing in the OSBL is due to the weakened modeled vertical shear of horizontal currents and buoyancy gradients resulting from a larger d$z$. 

Overall, KPP simulations with the NM configuration are similar to the base configuration. Adding diffusivities from interior mixing schemes (e.g., shear instability) can slightly reduce biases in a few cases (e.g. HW and CEW), but can increase temporal noise in the simulated OSBL.  Therefore, in the no-match configuration we do not recommend adding diffusivities from the shear instability scheme to the KPP diagnosed value in the no-match configuration.  

Finally we note that even though OSBL depths simulated by the no-match configuration are similar to the baseline configuration, the no-match configuration does not require the complexities associated with matching viscosities and diffusivities between the KPP scheme and internal mixing parameterizations.


\subsubsection{Considerations for the base configuration}

\begin{figure}[thbp]
	\centering\includegraphics[width=0.5\linewidth]{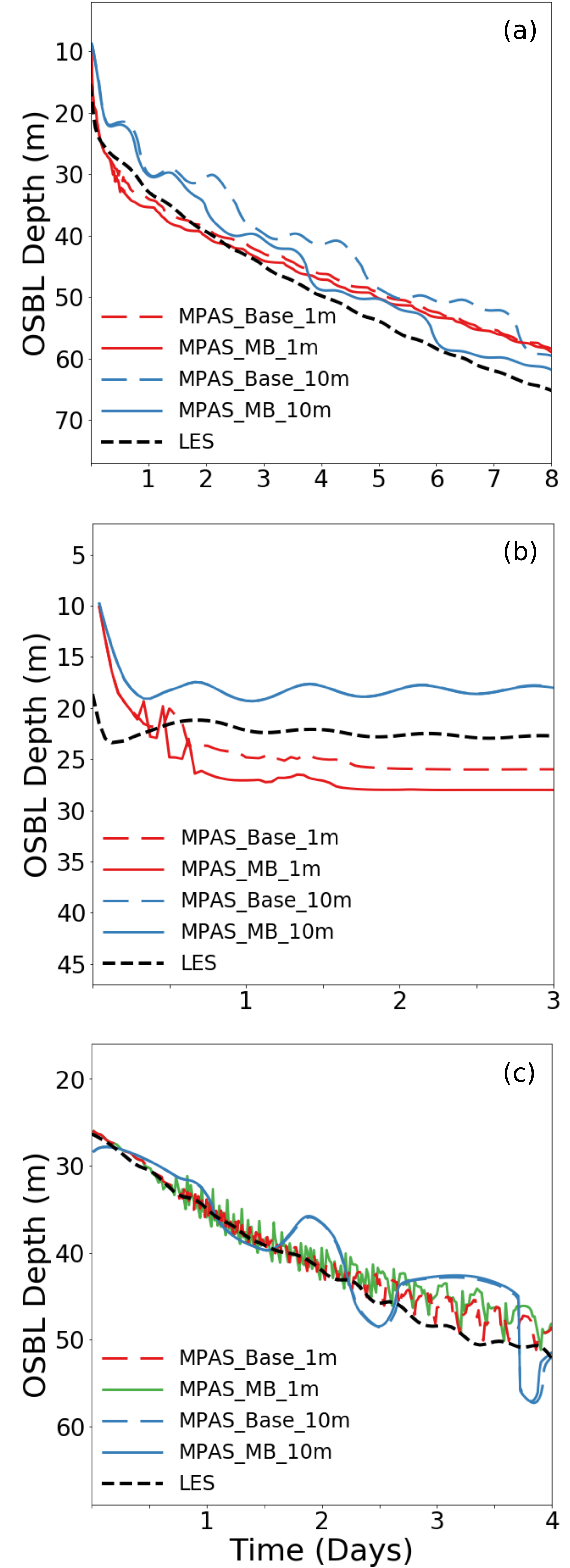}
	\caption{As in Figure~\ref{fig:NM_v_add} but for sensitivity tests matching the gradient of internal diffusivities.}
	\label{fig:mbTest}
\end{figure}

Recall that the CVMix base configuration of KPP does not match to gradients of internal diffusivities, which is different from LMD94 and \cite{Danabasoglu2006}.  Sensitivity to this choice is tested in Figure~\ref{fig:mbTest}.  In the CEW test there is minimal sensitivity to the matching of internal diffusivity gradients.  However, in the HW test (Figure~\ref{fig:mbTest}b), matching to internal diffusivity gradients increases the OSBL bias relative to LES at high resolution.

\begin{figure}[hbtp]
\centering\includegraphics[width=0.7\linewidth]{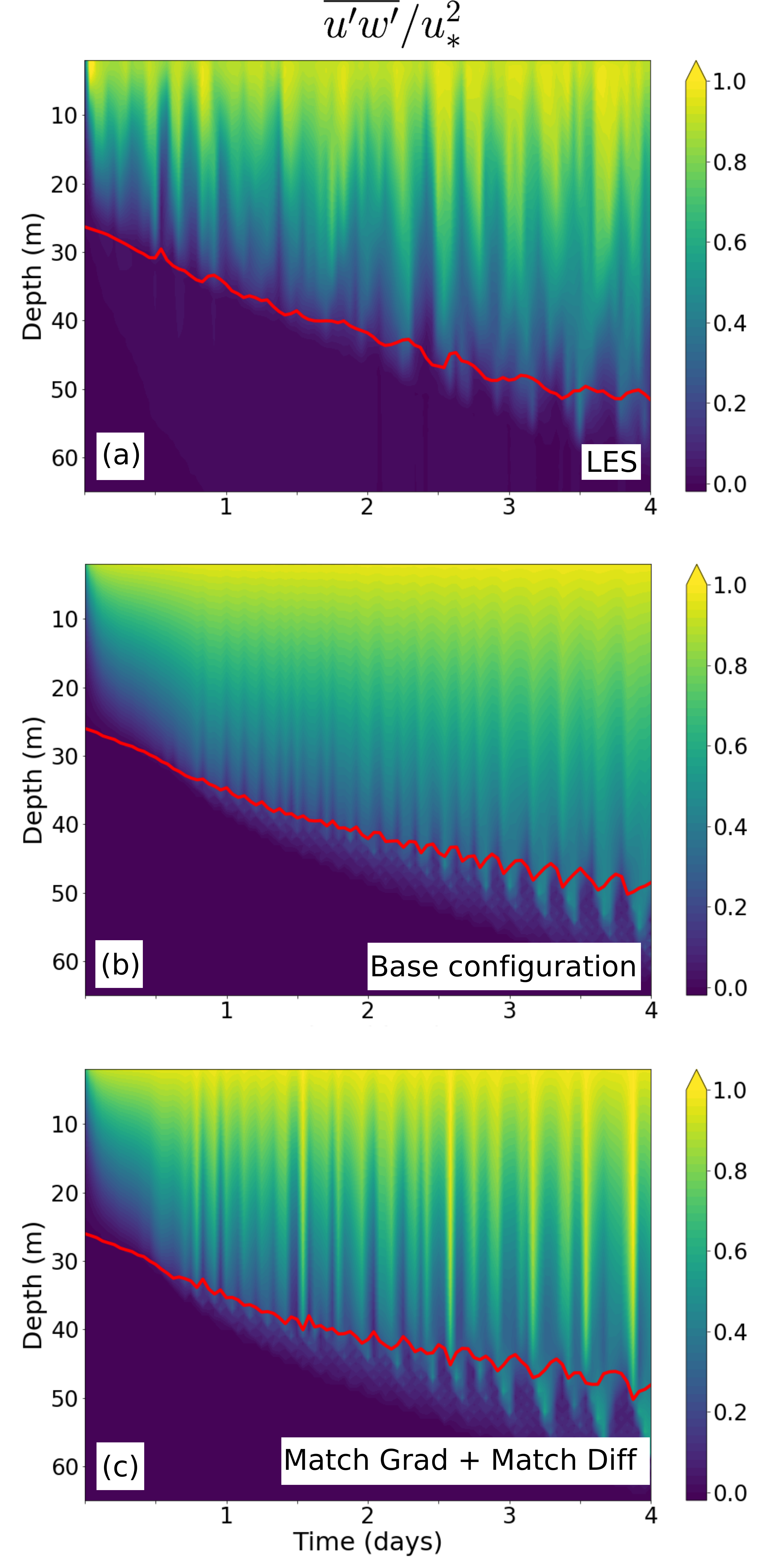}
\caption{Test Case CWB: Turbulent momentum flux normalized by the surface friction velocity (a-c).  The LES results are the top row, the middle row is the base configuration and the bottom row is a test where matching is to the internal diffusivity alone.  For KPP simulations, the turbulent fluxes are computed via the parameterizations in Equations~(\ref{KPPfluxText}),~(\ref{KPPDiffusivityText}), and~(\ref{non-localFlux}).  In every plot, the red line is the boundary layer depth computed via Equation (\ref{eq:KPP-boundary-layer-depth-defined}).}
\label{wind2_fluxes_matching}
\end{figure}

In the CWB test, the MB (matching diffusivity and the gradient from internal mixing) test result in brief periods of anomalously strong momentum fluxes (\textit{e.g.}, near day 3 in Figure~\ref{wind2_fluxes_matching}b).  When the OSBL deepens into the layer of strong background shear, the interior predicted diffusivities can vary rapidly near the OSBL base and thus strongly influence the KPP OSBL diffusivities.  When we match to the interior predicted diffusivities and viscosities alone (CVMix default), these regions of vigorous fluxes are strongly damped (compare Figures~\ref{wind2_fluxes_matching}b and c).

Our results (in particular the CWB test) suggest that if internal matching is used, matching should be to internal diffusivities only and not the gradient.

\subsection{Property exchange across the OSBL base}
\label{section:PEresults}
KPP models entrainment buoyancy fluxes via:
\begin{itemize}
\item the model-resolved vertical shear of horizontal currents
\item the unresolved turbulent shear ($V_t^2$) parameterization,
\item and the enhanced diffusivity parameterization.
\end{itemize}
The first two methods directly alter the boundary layer depth via equation~(\ref{eq:KPP-boundary-layer-depth-defined}).  As the resolved shear of horizontal currents and unresolved turbulent shear increase, the bulk Richardson number decreases, deepening the boundary layer.  A deeper boundary layer increases the entrainment buoyancy flux.

\begin{figure}
\centering\includegraphics[width=0.9\linewidth]{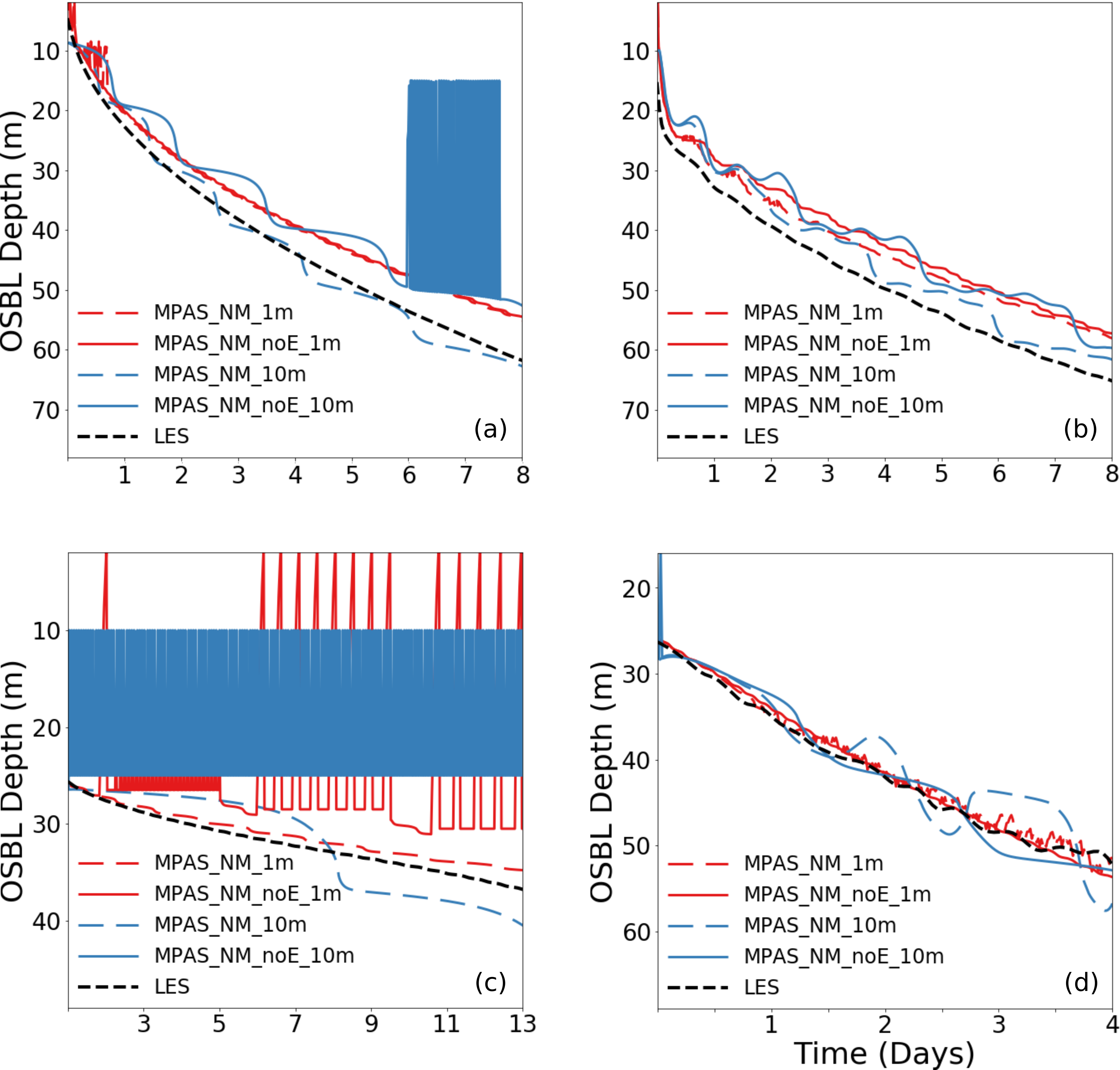}
\caption{Boundary layer depth sensitivity to the enhanced diffusivity parameterization in the no matching configuration.  The LES output is dashed black.  The baseline configuration results are reproduced as dashed lines.  All OSBL depths are computed using Equations~(\ref{eq:KPP-boundary-layer-depth-defined}). (a) The FC test, (b) the CEW test, (c) the FCML test, and (d) the CWB test.}
\label{fig:noEntrainmentBLD}
\end{figure}

LMD94 describe the enhanced diffusivity parameterization as a method to overcome a shallow bias and boundary layer "stair case" structures at coarse resolution.  This suggests that the enhanced diffusivity parameterization attempts to represent unresolved sources of entrainment.  Yet our tests show that the enhanced diffusivity parameterization can lead to strong resolution dependence in simulated boundary layer depths.  Figure~\ref{fig:noEntrainmentBLD} shows boundary layer depths for tests without the enhanced diffusivity parameterization.  In test cases with small vertical shear of resolved horizontal momentum (e.g. a-c) the coarse resolution OSBL depth shallows when the enhanced diffusivity parameterization is disabled.  Note that while the OSBL depth bias relative to LES increases, the KPP resolution dependence is diminished (Figure~\ref{fig:noEntrainmentBLD}a and b especially)

\begin{figure}
\centering\includegraphics[width=0.9\linewidth]{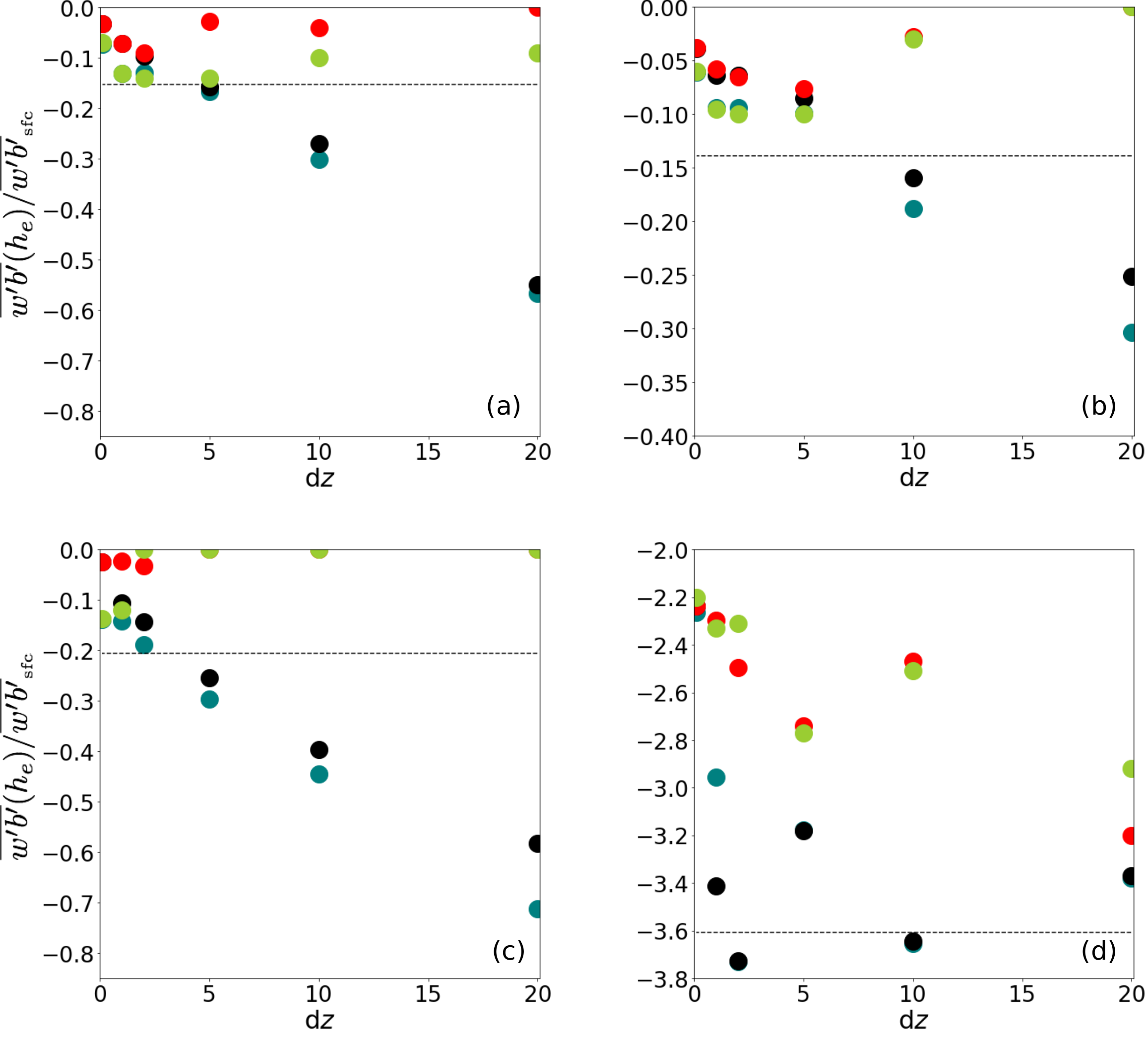}
\caption{Time averaged (over last day of simulation) entrainment buoyancy flux, which is defined as the minimum buoyancy flux in the profile, normalized by the surface buoyancy flux.  The entrainment buoyancy flux is shown as a function of KPP resolution.  The black circles include the enhanced diffusivity parameterization and the red circles do not.  The yellow-green circles include the alteration to $V_t^2$ but disables the enhanced diffusivity parameterization and the teal circles include both changes.  The LES entrainment buoyancy flux is shown as the horizontal dashed black line in each figure.  (a) The FC test, (b) the CEW test, (c) the FCML test, and (d) the CWB test.}
\label{fig:EntrainmentBuoyFlux}
\end{figure}

To examine the resolution dependent bias in more detail, Figure~\ref{fig:EntrainmentBuoyFlux} shows the time averaged entrainment flux across a wide range of vertical grid spacings (d$z=0.1$m - d$z=20$m) for four test cases (FC, CEW, FCML, and CWB).  In the FC, CEW, and FCML cases the entrainment buoyancy flux simulated by the base configuration is strongly dependent on resolution (black circles).  In the CWB test, there is minimal variation due to the large imposed background shear.

If the enhanced diffusivity parameterization is disabled, the entrainment buoyancy flux is strongly reduced in Figures~\ref{fig:EntrainmentBuoyFlux}a-c.  This further demonstrates that the enhanced diffusivity parameterization seeks to represent unresolved sources of entrainment.

The weak entrainment when enhanced diffusivity is disabled leads to large temporal noise in the FC and FCML test cases (Figures~\ref{fig:noEntrainmentBLD}a and c).  Recall that the non-local tracer flux (equation~(\ref{eq:gammab-defined})) causes cooling near the OSBL base in the presence of surface cooling via a redistribution of the surface flux.  Without sufficient entrainment, static instability can develop near the OSBL base.  When $N^2 < 0$, the unresolved turbulent shear is assumed to be zero.  This increases the bulk Richardson number (equation~(\ref{eq:KPP-boundary-layer-depth-defined})), which shallows the OSBL.

These large oscillations are evident in similar MOM6 and POP configurations as well (not shown).  While the mean deepening of the OSBL remains consistent in many cases even without sufficient entrainment, we note that the rapidly oscillating OSBL depths could have strong interactions with other parameterizations (\textit{e.g.}, symmetric instability, \citealp{bachman2017parameterization}).  

If the definition of $N_{osbl}$ is altered (equation~(\ref{eq:newN-defn})), the entrainment buoyancy flux increases in the base configuration (teal circles, Figure~\ref{fig:EntrainmentBuoyFlux}) as expected.  If the enhanced diffusivity parameterization is disabled in addition to altering $N_{osbl}$ (yellow-green circles, Figure~\ref{fig:EntrainmentBuoyFlux}) the resolution dependence is reduced, although the entrainment flux is too weak.  These results suggest that the enhanced diffusivity parameterization leads to the observed resolution dependence for KPP simulated OSBL depths (Figure~\ref{matchconfig_compare}).  However, the enhanced diffusivity parameterization is necessary to provide sufficient entrainment in order to prevent temporal noise in OSBL depths.  We discuss this issue further in light of the non-local tracer transport parameterization in the next section.

\section{Non-local transport in KPP}
\label{section:NLresults}
We focus in this section on the parameterized non-local tracer transport from KPP. The non-local transport is enabled only for destabilizing surface buoyancy forcing.  Across most test cases, the KPP non-local transport parameterization behaves well.  Yet our experiments also suggest possibilities for extensions worthy of further research. 

\subsection{Tracer transport}
\label{section:results-tracertrans}

FC sensitivity tests show that if entrainment is not predicted correctly (\textit{i.e.}, too weak), the OSBL oscillates rapidly due to an interaction with the non-local temperature flux parameterization.  This bias is mitigated by the enhanced diffusivity parameterization.  However, for shallow OSBL depths at high resolution, temporal noise is evident even with the enhanced diffusivity parameterization enabled (Figure~\ref{matchconfig_compare}a).

In the DC test we examined the sensitivity of the non-local temperature flux (equation (\ref{eq:gammab-defined})) to the amount of shortwave radiation included in the surface flux.  We varied the amount of shortwave radiation in the surface flux from that absorbed in the top layer to that absorbed in the entire OSBL.  Across these tests, we found very little variation in the simulated OSBL depths (not shown).  This is consistent with LMD94 (see their Figure C4, Appendix C) as this study and LMD94 use an equation of state where density increases until freezing occurs (\textit{e.g.}, equation~(\ref{eq:buoyancy})).  

In KPP, the non-local tracer transport derives from the buoyant production term in the turbulent flux equation (e.g. equation~(\ref{temperatureFlux})) and is only non-zero in the presence of a non-zero surface tracer flux.  In the FCML test there is no non-local salinity transport even though there is a destabilizing buoyancy flux.

\begin{figure}[hbtp]
\centering\includegraphics[width=0.99\linewidth]{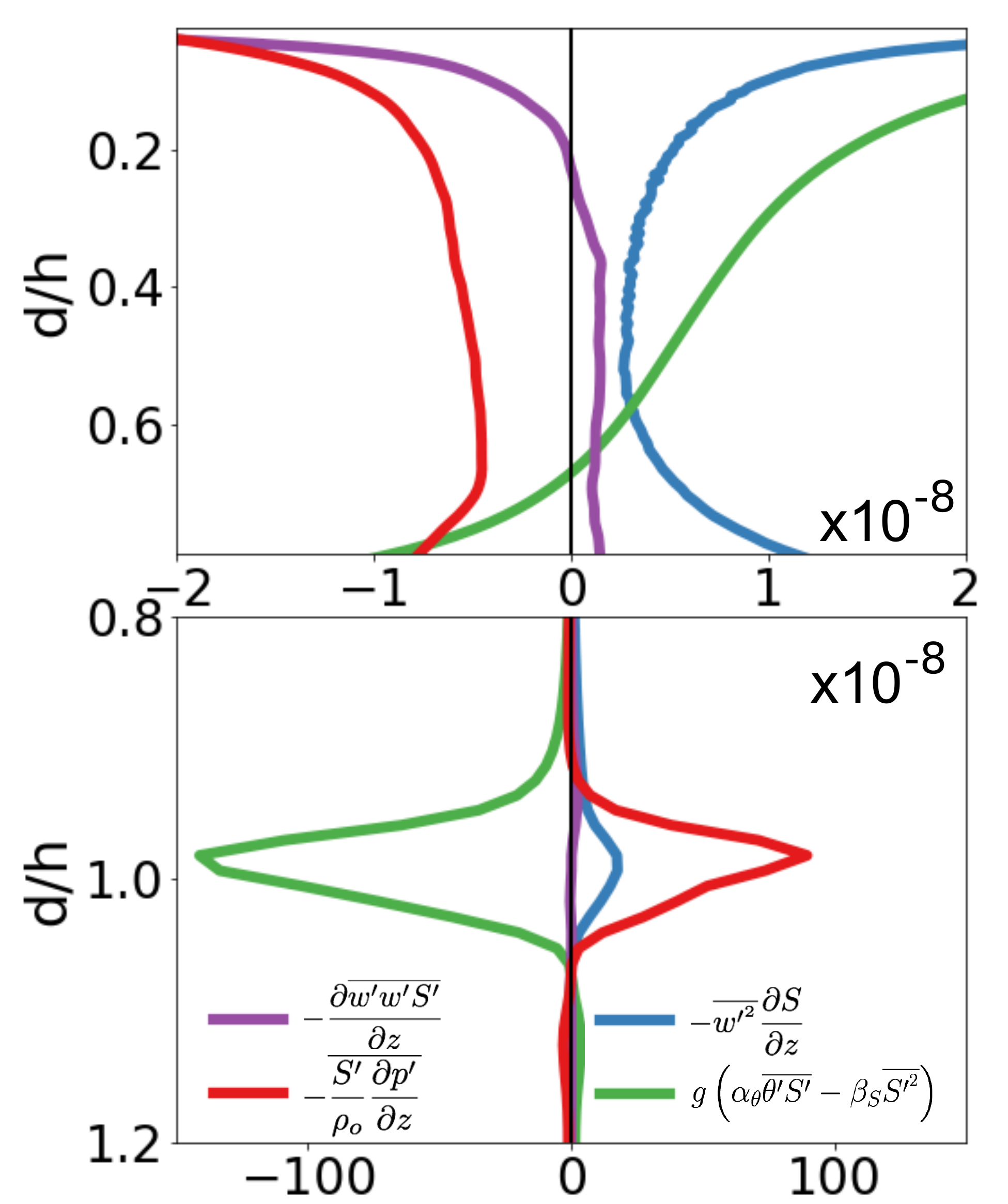}
\caption{Test Case FCML: budget terms for the turbulent salinity flux equation (see Equation~(\ref{salinityFlux})) diagnosed from LES.  The red line is the pressure-salinity covariance, the green line is the buoyant production term, the blue line is the local production term, and the purple line is the triple moment term.  All data has been averaged over the final 12 hours of day 12.}
\label{convect3_salinityFluxBudget}
\end{figure}

To examine this assumption, we show the salinity flux budget tendency terms in equation~(\ref{salinityFlux}) from the FCML LES test case in Figure~\ref{convect3_salinityFluxBudget}, where the two panels span the entire boundary layer, but the x-range in the upper OSBL is tightened to better elucidate the profiles.  Note that the sum of the tendency terms in Figure~\ref{convect3_salinityFluxBudget} is not exactly zero, most likely due to the lack of inclusion of subgridscale turbulent flux tendencies in Equation~(\ref{salinityFlux}) \citep{Mironov:2001gh}. 
\begin{equation}
\label{salinityFlux}
\frac{\partial \overline{w^\prime S^\prime}}{\partial t} =  -\underbrace{\overline{w^{\prime^2}} \frac{\partial \overline{S}}{\partial z}}_\textrm{local} + \underbrace{g \left(\alpha_\theta \overline{\theta^\prime S^\prime} - \beta_S \overline{S^{\prime^2}}\right)}_\textrm{buoyancy} - \underbrace{\frac{\partial \overline{w^\prime w^\prime S^\prime}}{\partial z}}_\textrm{triple moment} - \underbrace{\overline{\frac{S^\prime}{\rho_o}\frac{\partial p^\prime}{\partial z}}}_{\substack{\textrm{pressure-salinity}\\
\textrm{covariance}}}.
\end{equation}
Figure~\ref{convect3_salinityFluxBudget} shows that the buoyant production of the turbulent salinity flux (recall the non-local transport in KPP derives from this term) is larger in magnitude to the local production of the turbulent salinity flux in the OSBL.  The non-zero non-local production of the turbulent salinity flux in the FCML test implies that the KPP formulation of non-local tracer transport is incomplete.  Indeed, \cite{Noh2003} show that inclusion of a non-local redistribution of the entrainment heat flux results in a reduced bias for an atmospheric vertical mixing parameterization similar to KPP relative to an atmospheric LES. A parameterization similar to \cite{Noh2003} could replace the enhanced diffusivity parameterization of LMD94 and perhaps reduce the resolution dependence seen in predicted OSBL depths given there is no explicit consideration of grid resolution in the \cite{Noh2003} parameterization.

\subsection{Momentum transport}
\label{section:results-momentumtrans}

\begin{figure}[hbtp]
\centering\includegraphics[width=1\linewidth]{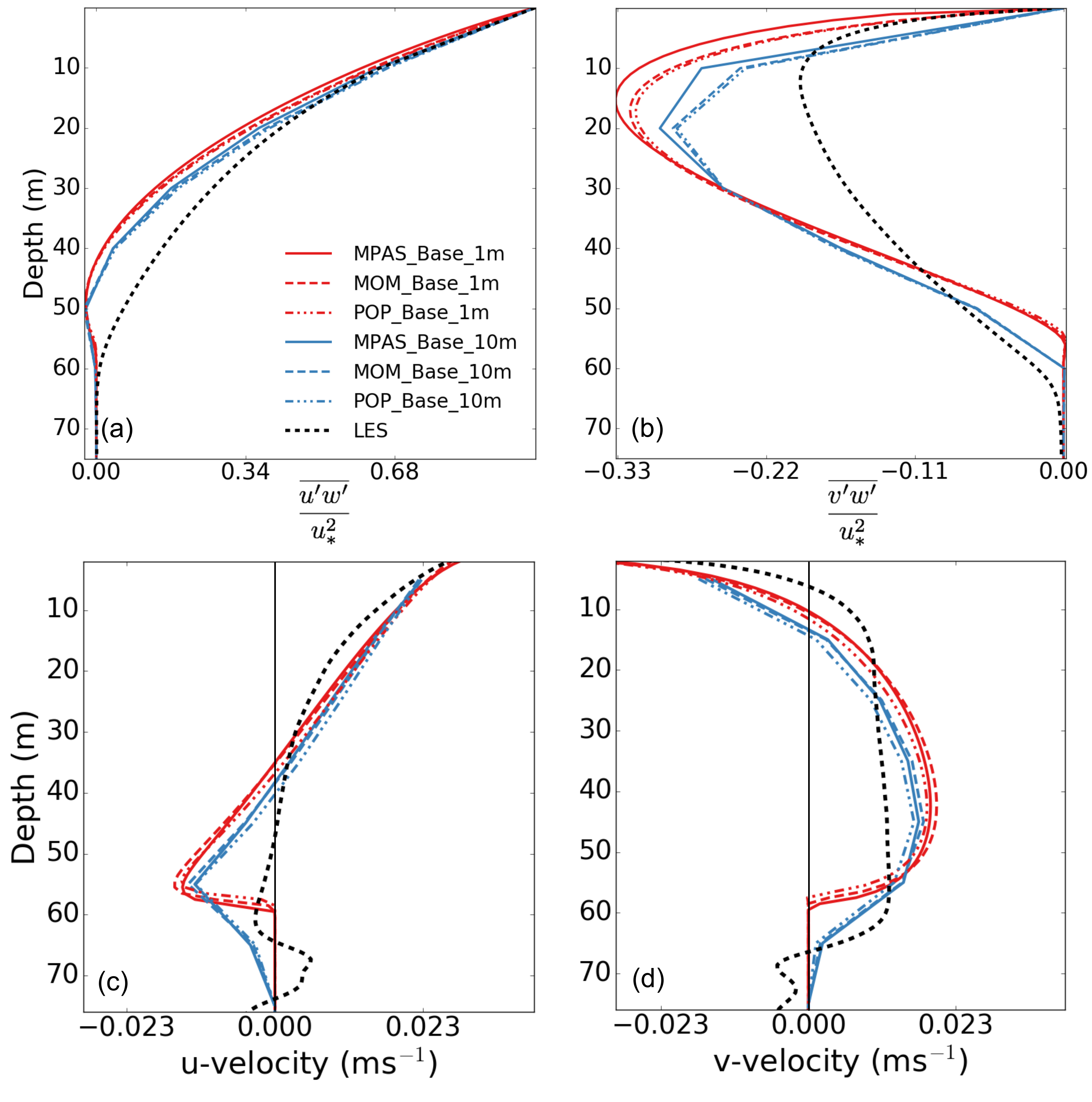}
\caption{Test Case CEW: time averaged, normalized momentum flux: (a) zonal momentum flux ($\overline{u^\prime w^\prime}$), (b) meridional momentum flux ($\overline{v^\prime w^\prime}$), (c) zonal velocity, and (d) meridional velocity.  The KPP momentum fluxes are computed via the parameterizations in Equations~(\ref{KPPfluxText}),~(\ref{KPPDiffusivityText}), and~(\ref{non-localFlux}).  The line colors are defined in Figure~\ref{matchconfig_compare}.  All profiles have been averaged over day 8.}
\label{convect2_momentumFlux}
\end{figure}

Figure~\ref{convect2_momentumFlux} shows the zonal and meridional momentum fluxes for the CEW test.  The base configuration is shown for MPAS-O, MOM6, and POP.  In every model, the KPP simulated zonal momentum flux matches LES fairly well (Figure~\ref{convect2_momentumFlux}a).  However, the KPP predicted meridional momentum flux profile is much too strong through much of the OSBL (Figure~\ref{convect2_momentumFlux}b).  The vigorous LES meridional momentum flux in the presence of minimal velocity gradient (Figure~\ref{convect2_momentumFlux}d) suggests an important role for non-local momentum transport.  The lack of a non-local momentum transport could explain the OSBL shallow bias seen in the CEW test (Figure~\ref{matchconfig_compare}b).

\section{Conclusions and recommendations}
\label{discussion}

In this paper, we investigated the behavior of the KPP boundary layer parameterization as implemented within the CVMix package \citep{Griffies2015}. The KPP scheme was rigorously tested against a series of horizontally averaged LES.  Further, in each case, we also varied $\mbox{Ri}_{\mbox{\tiny crit}}$.  When $\mbox{Ri}_{\mbox{\tiny crit}}$ was increased, the boundary layer did deepen, but not linearly due to $V_t^2$.  

While many different implementations of KPP have been tested in a number of circumstances; \textit{e.g.}, observations \citep[(\textit{e.g.}, LMD94,][]{Zedler2002,VanRoekel2012,mukherjee2016}, large-scale ocean simulations \citep[\textit{e.g.},][]{Li2001}, and against limited LES \citep[\textit{e.g.},][]{Large1999, McWilliams2000, Smyth2002, noh2016parameterization,Reichl:2016db}, it has not been subject to testing against a series of tightly controlled LES cases.  These tests focused on implementation choices related to vertical physics alone under simplified surface forcing.  The tests presented here do not consider the influence of horizontal processes. This is an important and active \citep{bachman2017parameterization} direction for future evaluation of KPP.

Our tests have focused on four main components of KPP physics and numerical choices: vertical resolution and timestepping (Sections~\ref{section:modelchoices} and~\ref{section:RTresults}), internal matching (Sections~\ref{sec:shapefunc-diffusmatch} and~\ref{section:DMresults}), entrainment at the OSBL base (Sections~\ref{section:entrainment} and~\ref{section:PEresults}), and non-local transport (Sections~\ref{subsection:non-local-transport} and~\ref{section:NLresults}).  The key findings from the most critical areas and a few potential solutions are summarized below
\begin{itemize}
\item \textsc{Model Choices}: 
\begin{itemize}
\item The CVMix implementation appears robust across calling models as three very different calling models gave similar results (Section~\ref{section:RTresults}).
\item KPP is sensitive to the vertical resolution chosen (Section~\ref{section:vertical-resolution}).  Using Equation (\ref{eq:newN-defn}) in the $V_t^2$ parameterization mitigates some of this resolution-dependent bias (\textit{e.g.}, Figure~\ref{n2_osbl_comparison} and~\ref{fig:EntrainmentBuoyFlux}).
\item KPP is robust to variation in model timestep (Figure~\ref{wind1_bld}).
\end{itemize} 
\item \textsc{Non-local transport}:
\begin{itemize}
\item The results of the FCML test case suggest that a non-local transport based on surface fluxes alone is not sufficient (Figure~\ref{convect3_salinityFluxBudget}) and a non-local redistribution of the entrainment buoyancy flux should be explored.
\item Results of the CEW and CWB tests (Figure~\ref{convect2_momentumFlux}) suggest that a non-local momentum transport parameterization is a critical need for KPP (Section~\ref{sec:futureNLT}).
\end{itemize}
\item \textsc{Treatment at the Boundary Layer Base}:
\begin{itemize}
\item It has been suggested (LMD94) that $\mbox{Ri}_{\mbox{\tiny crit}}$ should vary with vertical resolution and previously proposed functions of $\mbox{Ri}_{\mbox{\tiny crit}}(dz)$ were found by matching to d$z = 1$m results.  However, determining the correct form of $\mbox{Ri}_{\mbox{\tiny crit}}(dz)$ is further complicated for models with non-uniform vertical resolution (\textit{e.g.}, MPAS-O).  For simplicity, it is recommended to choose a single value of $\mbox{Ri}_{\mbox{\tiny crit}}$ across most resolutions. 
\item KPP simulates entrainment buoyancy fluxes in two ways.  At high resolution, the entrainment fluxes are very sensitive to the $V_t^2$ parameterization. Use of Equation (\ref{eq:newN-defn}) in the $V_t^2$ parameterization reduces this sensitivity. At coarse resolution, entrainment strength is determined by the enhanced diffusivity parameterization.  In a number of tests this parameterization caused a bias that grew with resolution (Figure~\ref{n2_osbl_comparison}).
\item For shallow boundary layers and at high resolution (\textit{e.g.}, the DC test case), high frequency OSBL depth noise develops (Figure~\ref{matchconfig_compare}c). This suggests a lack of simulated entrainment within KPP for this test case.
\item KPP OSBL biases relative to LES are similar for configurations with and without matching to interior diffusivities (compare Figures~\ref{matchconfig_compare}b and d to Figure~\ref{nomatch_coparison}).  Our results suggest considerations for both configurations of KPP (base and no match) that are summarized below:
\begin{itemize}
\item \textit{Matching to interior mixing:} Matching to the internal diffusivity and its vertical gradient can lead to noise in the simulated OSBL depths (Figure~\ref{fig:mbTest}) and periods of anomalously strong fluxes (Figure~\ref{wind2_fluxes_matching}.  Therefore, if matching is retained we recommend matching to the internal diffusivity only and \textbf{not} its gradient.
\item \textit{No matching:} diffusivities and viscosities from the LMD94 shear instability parameterization should \textbf{not} be considered in the OSBL.  Including interior diffusivities in the OSBL reduces the OSBL depth bias slightly in two test cases (CEW and HW, Figures~\ref{fig:NM_v_add}a-b), but introduces temporal oscillations in the OSBL depth in the CWB test case (Figure~\ref{fig:NM_v_add}c)
\end{itemize}
\end{itemize}
\end{itemize}

\section{Future work}
\label{section:futureWork}
In addition to proposed modifications to the baseline configuration KPP and a new configuration that does not match to internal predicted diffusivities/viscosities, the results also suggest directions for future improvements to KPP.

\subsection{Non-local transport}
\label{sec:futureNLT}

Results of the FCML test case suggest that assuming the non-local flux redistributes the surface flux alone is not complete.  There is an equally large contribution from non-local processes throughout the OSBL in the absence of surface fluxes in the FCML test (see Figure~\ref{convect3_salinityFluxBudget}).  Thus the current implementation for non-local tracer fluxes in KPP must be extended. Including a separate parameterization for non-local entrainment fluxes (similar to \citealp{Noh2003}) may be a viable path forward.

In the CEW test, the meridional momentum profiles in LES were well mixed in the middle of the OSBL, whereas in KPP a strong gradient exists (Figure~\ref{convect2_momentumFlux}d).  The meridional momentum flux for LES and KPP is strong in the OSBL (Figure~\ref{convect2_momentumFlux}c).  The presence of a meridional momentum flux in the presence of minimal vertical gradients of meridional momentum suggest an importance of non-local momentum transport.  

We see two possible paths for the inclusion of non-local momentum transport in KPP. First, we could use the form suggested by \cite{Smyth2002}.  To the best of our knowledge, this parameterization has not been tested outside of \cite{Smyth2002} even though it is based on some theoretical considerations and a similar parameterization has been tested in atmospheric LES \citep{frech1995two,brown1997non}.  Thus the \cite{Smyth2002} parameterization is not yet recommended for inclusion in CVMix.

Alternatively, we could derive an explicit equation for the momentum flux following \cite{DONALDSON1972}.  To make the equation tractable, we could also assume that any product of perturbations involving horizontal quantities (\textit{e.g.}, $u^\prime$) alone and products involving the Coriolis parameter ($f$) are negligible.  If we use the pressure strain correlation closure of \cite{canuto2007non}, we can write an equation for the turbulent vertical flux of zonal momentum as, 
\begin{equation}
\label{eq:non-local_momentum}
\overline{u^\prime w^\prime} \approx -\tau \overline{w^{\prime^2}} \left(\frac{\partial \overline{U}}{\partial z} -\frac{1}{\overline{w^{\prime^2}}} \frac{\partial \overline{u^{\prime} w^{\prime^2}}}{\partial z} - \frac{C_1 \overline{u^\prime b^\prime}}{\overline{w^{\prime^2}}} \right).
\end{equation}
Here, $\overline{u^\prime b^\prime}$ is the horizontal buoyancy flux, and $C_1$ is a constant.  If we ignored the horizontal buoyancy flux, we could perhaps follow HM91 to derive a closure for non-local momentum transport.  Yet, \cite{Holtslag1991} considered free convective conditions and thus the potential application to a non-local momentum parameterization is tenuous.  Additional LES cases must be considered to determine how to appropriately parameterize the second and third terms in Equation~(\ref{eq:non-local_momentum}).

Within a possible non-local momentum transport it is important to include enhancement due to Langmuir turbulence.  Previous work \citep{McWilliams2000,Li:2015gv,Reichl:2016db} have suggested parameterizations for the influence of Langmuir turbulence on the local diffusivity in KPP, but a non-local momentum parameterization that includes Langmuir turbulence has not been proposed.  It is possible that previously derived vertical velocity scalings \citep[\textit{e.g.}, ][]{McWilliams2000,van2012form} could be used to derive a modified convective velocity scale for a new non-local momentum flux parameterization following the equation~(\ref{eq:non-local_momentum}) or to modify the non-local momentum flux proposed by \cite{Smyth2002} that includes Langmuir turbulence.

\subsection{Amount of incident shortwave radiation transported by the non-local parameterization}
\label{shortWaveExperiment}
As suggested by LMD94, the upper limit for the shortwave contribution to the non-local vertical heat flux is the amount of radiation absorbed in the OSBL\footnote{The amount of shortwave radiation to include in the surface buoyancy forcing is a choice left to the calling model.  CVMix simply returns a non-dimensional non-local transport and the calling model scales this by the appropriate surface tracer flux.}.  A simple gedanken experiment can demonstrate a pitfall to choosing this upper bound in models with penetrating shortwave radiation.  Assume brackish surface waters, with a SST below the temperature of maximum density.  Further assume that local vertical mixing is small and shortwave radiation is the only heat flux.  Finally assume a deep OSBL.  In this circumstance solar heating leads to an unstable OSBL, and thus the non-local parameterization is active.  With a deep OSBL, all of the incident shortwave radiation is included in the non-local tracer transport, which results in a solar heat flux and non-local heat flux that exceeds the surface heat flux.  Thus new cold water masses would be generated near the surface that were not present in the water column.  In a more realistic configuration, this could lead to the artificial generation of sea ice.

A possible solution is to make the strength of the non-local flux depth dependent and to only include the portion of the shortwave radiation that is absorbed between the sea surface and a given depth in the non-local temperature transport.  This ensures that the redistributed heat flux is smaller than the total shortwave radiation entering the ocean.  Testing this solution and determining the appropriate amount of shortwave radiation to include in the non-local tracer transport would require modifications to the buoyancy equation and subgrid TKE schemes in the LES model.

\subsection{OSBL entrainment}
\label{sec:futureEntrainment}

To reduce the dependence of the entrainment flux across resolutions and forcing we could follow the suggestion of \cite{Noh2003}, who suggest an explicit parameterization for the entrainment heat flux, including a non-local redistribution of this flux.  In KPP the vertical entrainment heat flux is artificially separated from the remainder of the OSBL turbulent flux (and slaved to the OSBL depth prediction).  A parameterization for an entrainment heat flux similar to \cite{Noh2003} would make the boundary layer depth a diagnostic variable with no feedbacks into the scheme.  A diagnostic OSBL depth could eliminate mixing sensitivity to $V_t^2$ and the enhanced diffusivity parameterization in KPP.  

In summary, the current configuration of KPP has performed well in many scenarios.  However,  alternative approaches or extensions to KPP are needed to improve model fidelity for several relevant dynamic and thermodynamic situations.   Research into these challenging areas is underway.

\acknowledgments
We thank Scott Bachman, Mark Petersen, Rachel Robey, Milena Veneziani, and Phillip Wolfram for providing very useful comments on earlier drafts of this paper.  We also thank Mathew Maltrud for help in configuring POP simulations.  This research was supported as part of the Energy Exascale Earth System Model (E3SM) project, funded by the U.S. Department of Energy, Office of Science, Office of Biological and Environmental Research.  Michael Levy and Brian Kauffman were partially supported by the U.S. Department of Energy, Office of Science, Office of Biological and Environmental Research SciDAC grant DE-SC0012605 to NCAR. NCAR is sponsored by the National Science Foundation.  This research used resources of the Oak Ridge Leadership Computing Facility at the Oak Ridge National Laboratory, which is supported by the Office of Science of the U.S. Department of Energy under Contract No. DE-AC05-00OR22725 through an ALCC allocation and resources provided by the Los Alamos National Laboratory Institutional Computing Program, which is supported by the U.S. Department of Energy National Nuclear Security Administration under Contract No. DE-AC52-06NA25396.

\appendix
\newpage 

\section{Elements of the KPP boundary layer scheme}
\label{sec:KPP-describe}

For any prognostic scalar or vector field component $\psi$ (\textit{e.g.}, velocity components, tracer concentrations), the KPP scheme parameterizes the turbulent vertical flux within the surface boundary layer according to 
\begin{equation}
\label{KPPflux}
\overline{w^\prime \psi^\prime} = -K_\psi \left(\frac{\partial \psi}{\partial z} - \gamma_\psi\right).
\end{equation}
In this equation, the eddy diffusivity $K_\psi$ is written as the product of three terms
\begin{equation}
\label{KPPDiffusivity}
K_\psi = h\, w_\psi \left(\sigma\right)G\left(\sigma \right).
\end{equation}
The boundary layer depth $h >0$ scales the diffusivity, so that $K_\psi$ is larger for deeper boundary layers. The non-dimensional shape function, $G(\sigma)$, is described in \ref{subsection:non-dimensional-shape-function}.  

The turbulent velocity scale, $w_\psi \ge 0$, is computed according to \begin{equation}
\label{wpsi}
w_\psi = \left( \frac{\kappa \, u_*}{\phi_\psi (\sigma h/L)} \right).
\end{equation}
In this expression, $\kappa = 0.4$ is the von K\'{a}rm\'{a}n constant, $u_* \ge 0$ is the friction velocity scale (determined by the square root of the surface stress magnitude), $\sigma$ is the non-dimensional boundary layer coordinate (see equation~(\ref{sigDefn})), and $L$ is the Obukhov length scale, which is held fixed at its surface value. The function $\phi_\psi \ge 0$ is a non-dimensional flux profile that is smaller for negative buoyancy forcing and goes to unity in the absence of buoyancy forcing. Given this form, the velocity scale $w_\psi$ is larger for unstable surface boundary forcing (\textit{i.e.}, negative buoyancy forcing such as when removing heat or adding salt), as well as for stronger mechanical forcing (\textit{i.e.}, larger friction velocity scale as under strong wind forcing). 

The KPP vertical viscosity (used for frictional transfer of momentum in the ocean interior) is specified via a separate dimensionless flux profile through the Prandtl number
\begin{equation}
Pr = \left( \frac{K_v}{K_\psi} \right) 
= \left( \frac{\phi_\psi}{\phi_m} \right).
\end{equation}
See Appendix B of LMD94 as well as \cite{Griffies2015} for full details of the non-dimensional flux profile functions $\phi_\psi$ and $\phi_m$, as well as the Obukhov length scale $L$.

\subsection{The non-dimensional shape function} 
\label{subsection:non-dimensional-shape-function}
In the KPP diffusivity expression (equation~(\ref{KPPDiffusivity})), the non-dimensional shape function, $G(\sigma)$, is assumed to take a polynomial form proposed by \cite{OBrien1970}
\begin{equation}
\label{shape_func}
G(\sigma) = c_1 + c_2 \sigma + c_3 \sigma^2 + c_4 \sigma^3,
\end{equation}
where $c_1, c_2, c_3, c_4$ are constants to be specified by the following considerations. First, since the diffusivity, $K_\psi$, is assumed to go to zero at the ocean surface,
\begin{equation}
 c_1 = 0.
\end{equation}
 
Within the surface layer ($0 \le \sigma \le \epsilon$) (see Figure \ref{BL_schematics}), we can eliminate the gradient of $\psi$ in equation~(\ref{KPPflux}) using Monin-Obukhov similarity theory in the form  
\begin{equation}
\label{MO_sfc_scaling}
\frac{\partial \psi}{\partial z} = 
\left( 
\frac{ \overline{w^\prime \psi^\prime}_{\mbox{\tiny sfc}}}{\kappa \, z \, u_*} 
\right)
\phi_\psi,
\end{equation}
where $\overline{w^\prime \psi^\prime}_{\mbox{\tiny sfc}}$ is the turbulent boundary flux crossing the ocean surface (\textit{e.g.}, turbulent latent and sensible heat; turbulent tracer flux; turbulent momentum flux). If we combine equations~(\ref{KPPflux}) and~(\ref{KPPDiffusivity}) and assume positive surface buoyancy forcing so that the non-local term vanishes ($\gamma_\psi = 0$) we have 
\begin{equation}
\label{fluxEquation}
\overline{w^\prime \psi^\prime} = -h \, G \, w_\psi \left( \frac{\partial \psi}{\partial z}\right),
\end{equation}
which returns us to the KPP closure form assumed in equation~(\ref{KPPflux}). Now insert equation~(\ref{MO_sfc_scaling}) into~(\ref{fluxEquation}), and assume $G(\sigma) \approx \sigma (c_2 + c_3 \sigma)$ (valid in the surface layer where $\sigma \le \epsilon \ll 1$), to yield 
\begin{equation}
\label{psifluxEqn}
\overline{w^\prime \psi^\prime} = 
-\left( 
\frac{\overline{w^\prime \psi^\prime}_{\mbox{\tiny sfc}}}{\kappa \, z \, u_*} \right)
 \phi_\psi \, h \, \sigma \, 
 (c_2 + \sigma c_3) \, w_\psi. 
\end{equation}
Using equations~(\ref{wpsi}) and~(\ref{sigDefn}) brings equation~(\ref{psifluxEqn}) to the form
\begin{equation}
\label{sfclayerrelation}
\left( \frac{\overline{w^\prime \psi^\prime}}{\overline{w^\prime \psi^\prime}_{\mbox{\tiny sfc}}}
 \right)
 = c_2 + \sigma c_3.
\end{equation}
Now we assume a linear decrease of the turbulent flux within the surface layer (\textit{i.e.}, $\overline{w^\prime \psi^\prime}\,\epsilon = \beta \, \overline{w^\prime \psi^\prime}_{\mbox{\tiny sfc}}$ where $\beta$ is a constant), so that the surface flux at a position $\sigma$ within the surface layer is given by
\begin{equation}
\label{fluxinsfclayer}
\left( \frac{\overline{w^\prime \psi^\prime}}{\overline{w^\prime \psi^\prime}_{\mbox{\tiny sfc}}}
 \right)
 = 1 + \frac{\sigma}{\epsilon} (\beta -1)
 = c_2 + \sigma \, c_3.
\end{equation}
To be valid at $\sigma = 0$ requires
\begin{equation}
c_2 = 1.
\end{equation}

To determine the final two shape function coefficients, we require  matching across the base of the boundary layer, at $\sigma=1$.  Use of equation~(\ref{shape_func}) and its derivative at the boundary layer base leads to the following expressions
\begin{subequations}
\begin{align}
c_3 &= -2 + 3\, G(1) - \left( \frac{\partial G}{\partial \sigma} \right)_{\sigma=1} 
\label{eq:c3}
\\
c_4 &= 1 - 2\, G(1) + \left( \frac{\partial G}{\partial \sigma} \right)_{\sigma=1}.
\label{eq:c4}
\end{align}
\end{subequations} 
Thus, the shape function is dependent on the chosen boundary conditions at the base of the OSBL.  We next consider these boundary conditions.  

\subsection{Diffusivity matching for the shape function at the OSBL base}
\label{subsection:diffusivity-matching-shape-function}

LMD94 suggest that the diffusivity and viscosity predicted by KPP, as well as its vertical derivative, should match that predicted by the sum of all 
mixing parameterizations in the region below the boundary layer (the ocean interior). To ensure appropriate matching, the necessary inputs to equations~(\ref{eq:c3}) and~(\ref{eq:c4}) are given by 
\begin{subequations}
\begin{align}
\label{internalMatching1}
G(\sigma) &= \frac{K_\psi^{\mbox{\tiny INT}} \left(h \right)}{h \, w_\psi(\sigma)}
\\
\label{internalMatching2}
\frac{\partial G}{\partial \sigma} &= -\left[
 \frac{\partial_z K_\psi^{\mbox{\tiny INT}}  \left(h\right)}{w_\psi(\sigma)} + \frac{K_\psi^{\mbox{\tiny INT}}\left(h\right) \partial_{\sigma} w_\psi(\sigma)}{h\,w_\psi^2(\sigma)}
 \right], 
\end{align}
\end{subequations}
 where $\partial_z$ and $\partial_\sigma$ are the partial derivatives with respect to $z$ and $\sigma$ respectively, and we evaluate terms on the right hand side at the boundary layer base, $\sigma=1$. Without diffusivity matching, the shape function takes the relatively simple cubic form used by \cite{troen1986simple}
\begin{equation}
\label{simpleshapesEqn}
G\left(\sigma\right) = \sigma \left(1-\sigma\right)^2. 
\end{equation}

\section{Mathematical Symbols}
\label{math_symbols}
A summary of selected symbols used in this paper, along with their preferred units, are presented in Table~\ref{symbol_table}.
\begin{table}
	\def\arraystretch{1.15}
	\centering
	\begin{tabular}{ l  p{7cm}  l }
	 \hline
	 Symbol & Description & units \\
	 \hline
	 $\overline{w^\prime \theta^\prime}_{\mbox{\tiny sfc}}$ & surface temperature flux & $\mbox{m}~\mbox{s}^{-1}~\mbox{K}$   \\
	 $\overline{w^\prime S^\prime}_{\mbox{\tiny sfc}}$ & surface salinity flux & $\mbox{m}~\mbox{s}^{-1}~\mbox{ppt}$ \\
	 $\overline{w^\prime b^\prime}_{\mbox{\tiny sfc}}$ & surface buoyancy flux  & $\mbox{m}~\mbox{s}^{-1}~\mbox{m}~\mbox{s}^{-2}$ \\
	 $h$ & ocean boundary layer depth & $\mbox{m}$\\
	 $h_e$ & depth of the minimum buoyancy flux (entrainment depth) & $\mbox{m}$ \\
	 $h_m$ & depth of the well-mixed layer & $\mbox{m}$ \\
   	 $\sigma=(-z+\eta) / h$ & boundary layer coordinate & non-dimensional \\
   	 $\epsilon = 0.1$ & surface layer depth as a percentage of $h$ & non-dimensional \\
   	 $X_{\tiny \mbox{sl}}$ & surface layer average of $X$ & dimensions of $X$ \\
	 $G(\sigma)$ & shape function for diffusivity profile & non-dimensional \\
	 $w_x$ & turbulent velocity scale of quantity $x$ & $\mbox{m}~\mbox{s}^{-1}$ \\
	 $K_x$ & parameterized KPP eddy diffusivity for quantity $x$ & $\mbox{m}^{2}~\mbox{s}^{-1}$ \\
  	 $\mbox{Ri}_{\mbox{\tiny b}}$ & bulk Richardson number & non-dimensional \\
	 $\mbox{Ri}_{\mbox{\tiny crit}}$ & critical bulk Richardson number & non-dimensional \\
  	 $V_t^2$ & squared unresolved turbulent velocity (in $\mbox{Ri}_{\mbox{\tiny b}}$) & $\mbox{m}^{2}~\mbox{s}^{-2}$\\
	 $\gamma_x$ & non-local term & 
     $\mbox{tracer conc}~*~\mbox{m}^{-1}$ \\
	 $K_x \, \gamma_x$ & non-local flux & $\mbox{tracer conc}~*~\mbox{m}~\mbox{s}^{-1}$ \\
	 $C_*$ & strength of non-local term & non-dimensional \\
	 $\kappa$ & von K\'arm\'an constant & non-dimensional \\
	 $\alpha_{\theta}$ & thermal expansion coefficient & $\mbox{K}^{-1}$ \\
	 $\beta_S$ & haline expansion coefficient & $\mbox{ppt}^{-1}$ \\
	 $\overline{\theta}$ & Horizontal mean of temperature & $\mbox{K}$ \\
	 $\overline{S} $ & horizontal mean of salinity & $\mbox{ppt}$\\
	 $N^{2} \equiv \partial b/\partial z$ & squared buoyancy frequency 
     & $\mbox{s}^{-2}$ \\
	 $\tau$ & return to isotropy timescale & $\mbox{s}$\\
	 $g$ & gravitational acceleration & $\mbox{m}~\mbox{s}^{-2}$ \\
	 $f$ & Coriolis parameter & $\mbox{s}^{-1}$ \\
	  
	 \hline
	 
	 \end{tabular}
\caption{Symbols used in this paper, along with preferred units}
\label{symbol_table}
\end{table}

\section{Salinity testing in the NCAR LES}
\label{salinity:testing}

\begin{figure}[t]
	\centering\includegraphics[width=0.7\linewidth]{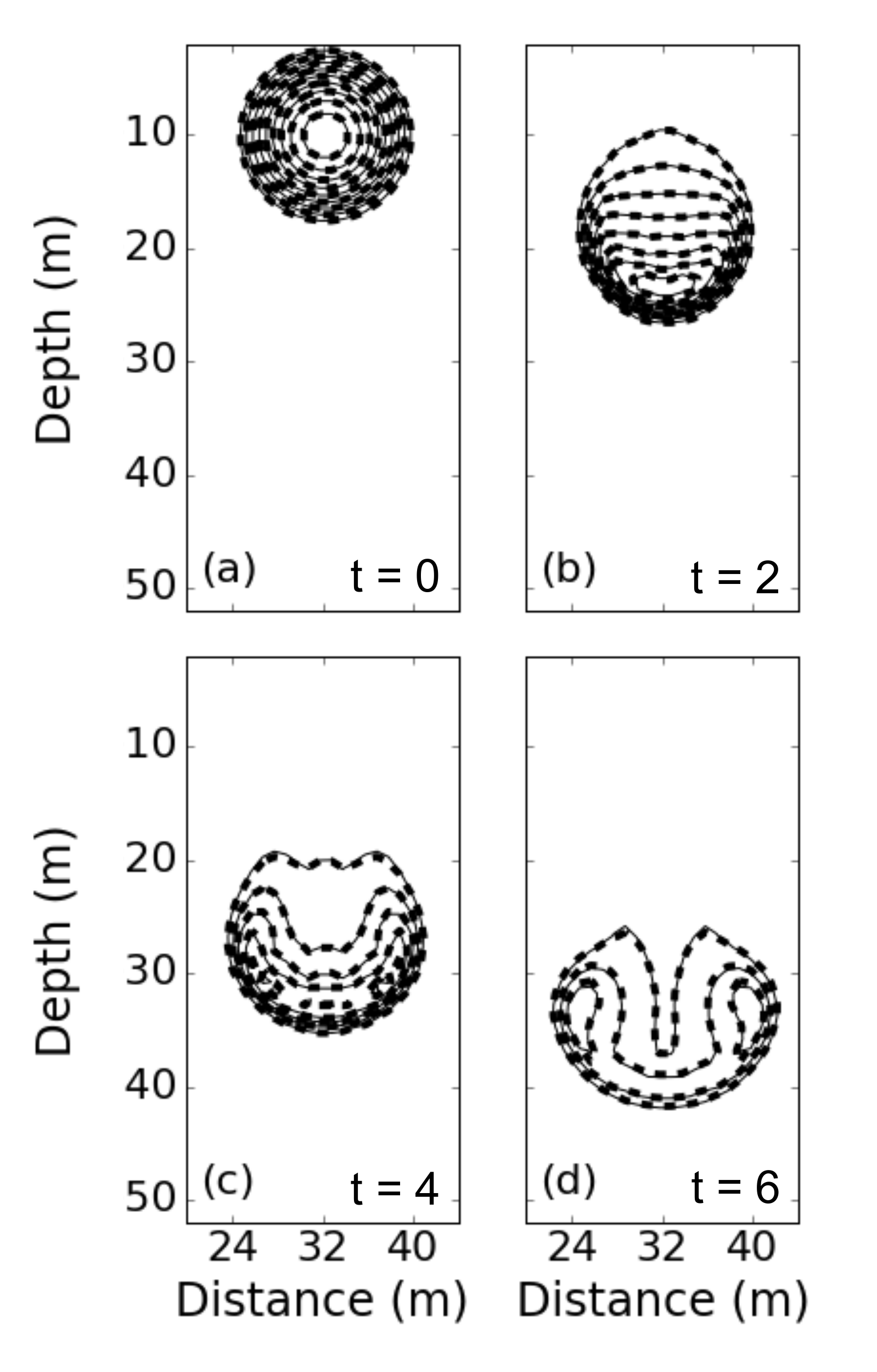}
	\caption{Results from a buoyant bubble test.  The thin solid lines are the buoyancy bubble dependent on temperature only.  The thicker dashed lines are results from the buoyant bubble dependent on salinity alone.  The respective perturbations in temperature and salinity have equivalent buoyancy perturbations. Labels in panels (a) - (d) are the time in minutes from the start of the simulation.}
	\label{saltBubble}
\end{figure}

To simulate the influence of salinity in the NCAR LES model \citep{McWilliams1997,sullivan2007surface}, the influence of salinity on the resolved buoyancy is included via a linear equation of state.  Thus the modified buoyancy term in the vertical momentum equation becomes
\begin{equation}
b = -g \left[1 - \alpha_{\theta} \left(\theta - \overline{\theta}\right)    + \beta_{S} \left(S - \overline{S}\right)\right]
\end{equation}
where the overline indicates a horizontal average.  Buoyancy terms in the subgrid TKE scheme are modified in a similar fashion.  To test this implementation two buoyant bubble tests were conducted.  In both tests, the buoyancy perturbation is initialized via
\begin{equation}
b(x,y,z) = min \left[0, \Delta b \; \cos\left(\frac{\pi \, r(x,y,z)}{2}\right) \right]
\end{equation}
where $\Delta b$ is the maximum buoyancy perturbation, and $r(x,y,z)$ is the distance from the bubble center.  One test is initialized by a temperature perturbation.  For the second test, the salinity is initialized such that in buoyancy the perturbation is identical to the temperature test.  The results from this test are shown in Figure~\ref{saltBubble}.  The thin solid lines are for the temperature perturbation and the thick dashed lines are the salinity perturbations.  As the two bubbles have identical buoyancy they should fall on top of each other.  Following from a-d in Figure~\ref{saltBubble} the bubbles stay together.

The buoyancy changes and the corresponding salinity flux implementation are validated in a free convection simulation due to surface evaporation (CEW; Table~\ref{exp_params}).  The agreement in the location of the buoyancy flux minimum diagnosed from the LES model and the analytic solution (see Appendix~\ref{analytic}) is quite good (not shown).  Given these results, we have confidence in the comparison of salinity from CVMix to LES for the simulations described in Tables~\ref{exp_params} - \ref{TS_profiles} and Sections~\ref{section:RTresults} - \ref{section:NLresults}.

\section{Form of solar radiation}
\label{shortwave:desc}
In the LES and single column tests, the shortwave radiation is assumed to obey a Jerlov type IB extinction profile \citep{Paulson1977}.  The time variation of incident shortwave radiation ($\mbox{W}~\mbox{}~\mbox{m}^{-2}$; $F_{sw}(t)$) is given as
\begin{equation}
\label{eq:Fsw}
F_{\mbox{\tiny sw}}(t) = Q_{\mbox{\tiny sw}}^{\mbox{\tiny max}} \max \left\{\cos \left[2 \pi \left(\frac{t}{T} - \frac{1}{2}\right)\right],0\right\}
\end{equation}
where $t$ is the time in seconds and $T=86400$ are the number of seconds in one day.  The peak incoming shortwave radiation ($Q_{\mbox{\tiny  sw}}^{\mbox{\tiny max}}$) is determined by forcing a balance between the daily integrated buoyancy gain and the daily integrated buoyancy loss ($\mbox{m}^2~\mbox{s}^{-3}$), \textit{i.e.},
\begin{equation}
\label{surface:solar}
\int_0^T \frac{g \, \alpha_{\theta}}{\rho \, c_p} \, F_{\mbox{\tiny sw}}(t) \, \mathrm{d}t \; 
= \; -\int_0^T \overline{w^\prime \, b^\prime}_{\mbox{\tiny sfc}}  \, \mathrm{d}t
\end{equation}
where this equation has been multiplied by the gravitational acceleration, $g$, and the thermal expansion coefficient, $\alpha_{\theta}$ (equation~(\ref{eq:alphaT})) and divided by the reference density and $c_p \equiv 4200 ~\mbox{J}~\mbox{kg}^{-1}~^{\circ}\mbox{C}^{-1}$ to ensure consistent units. Upon integration, and using equation~(\ref{eq:Fsw}), the maximum surface shortwave radiation is given as
\begin{equation}
\label{eq:qswMax}
Q_{\mbox{\tiny sw}}^{\mbox{\tiny max}} = 
-\pi \left( \frac{\rho \, c_p
\overline{w^\prime b^\prime}_{\mbox{\tiny sfc}} }  {g \, \alpha_{\theta}}
 \right).
\end{equation}
The units of equation~(\ref{eq:qswMax}) are ($\mbox{m}~\mbox{}^{\circ}\mbox{C}~\mbox{s}^{-1}$) given the division by $g \, \alpha_{\theta}$.  

\section{Analytic boundary layer solution}
\label{analytic}

The derivation in this appendix follows previous work closely \citep{turner1973bouyancy,haine1998gravitational}, but here we do not assume the entrainment of fluid into the boundary layer is negligible.  Assuming that entrainment into the boundary layer is negligible is equivalent to assuming that $h = h_e = h_m$ in KPP (Figure~\ref{Vt_schematics}).  In this derivation we relax this assumption slightly by assuming the buoyancy jump between the boundary layer and interior is large but not a discontinuity (\textit{i.e.}, $h \approx h_e \approx h_m$).

We begin by assuming that a model is forced with a horizontally uniform, constant in time, buoyancy flux, which simplifies the buoyancy equation to
\begin{equation} \label{buoy}
\frac{\partial b}{\partial t} = -\frac{\partial \overline{w'b'}}{\partial z}.
\end{equation}
Next, equation~(\ref{buoy}) is integrated from the surface $(z=0)$ to a depth $-H$ below the boundary layer ($h$) that does not change in time, \textit{i.e.}, 
\begin{equation} 
\label{int:0}
\int_{-H}^0 \frac{\partial b}{\partial t} \; dz = -\int_{-H}^0 \frac{\partial \overline{w'b'}}{\partial z} \; dz.
\end{equation}
Assuming $\overline{w'b'}$ is small at a depth $H$ below the boundary layer, equation~(\ref{int:0}) can be written as
\begin{equation} 
\label{int:1}
\frac{\partial}{\partial t} \int_{-H}^0 b(z) \; dz = -\overline{w'b'}_{z=0}.
\end{equation}

\begin{figure}[t]
\centering\includegraphics[width=.99\linewidth]{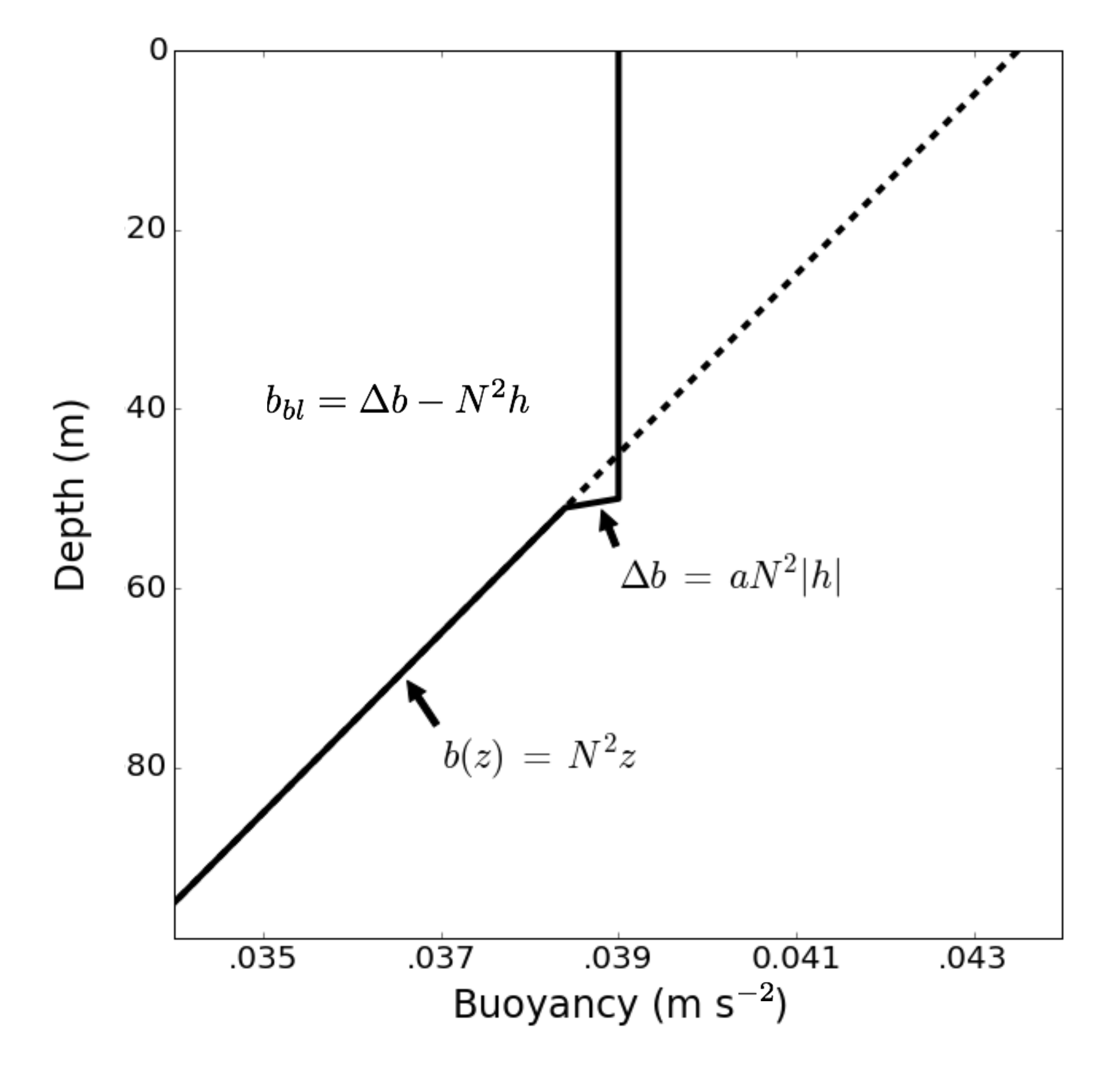}
\caption{Schematic illustrating the analytic buoyancy solution.  The dashed line is the initial stratification, and the solid line is the profile after some time (\textit{t}).  The buoyancy change across the entrainment layer is assumed to be a small fraction of the interior stratification ($a << 1$).  Thus this case does not assume a jump discontinuity between the well-mixed layer and the interior ocean.  Yet, given that the buoyancy change is large, $h_e \approx h$ (Figure~\ref{Vt_schematics}).}
\label{buoy_schematic}
\end{figure}

We now need to assume a buoyancy profile for the upper ocean.  The assumed form is shown in Figure~\ref{buoy_schematic}.  The buoyancy is uniform in the boundary layer, the stratification is constant below the boundary layer, and there is a sharp (but not discontinuous) buoyancy change between the well-mixed layer and the interior ocean, \citep[similar to][]{VanZanten1999}.  The strength of the buoyancy jump is assumed to be equal to the deep ocean buoyancy at the boundary layer base (\textit{i.e.}, $N^2 |h|$) multiplied by a small constant ($a$), which is assumed to be less than one.  With this assumption, the boundary layer buoyancy is 
\begin{equation}
\label{b_bl_defn}
b(h) = \Delta b - N^2 |h|= (a - 1) N^2 |h|.
\end{equation}
Using equation~(\ref{b_bl_defn}),~(\ref{int:1}) can be written as
\begin{equation}
\label{int:2}
\frac{\partial}{\partial t} \left[\int_{-H}^{-h(t)} N^2 z dz + \int_{-h(t)}^0 (a-1)N^2 |h| dz \right] = \overline{w^\prime b^\prime}_{\mbox{\tiny sfc}}.
\end{equation}
Equation~(\ref{int:2}) can be easily integrated to find
\begin{equation}
\label{apEq:3}
(1-2a)N^2\frac{\partial}{\partial t}\left[\frac{h^2}{2}\right]=\overline{w^\prime b^\prime}_{\mbox{\tiny sfc}}.
\end{equation}
The constant $a$ is determined by using the formula for the boundary layer heat flux from \cite{Lilly1968} and the empirical rule of convection 
\begin{equation}
 \frac{ \overline{w^\prime b^\prime}(h_e)}{\overline{w^\prime b^\prime}_{sfc}}  \approx 0.2.
\end{equation}
 Since the buoyancy change is sharp, $h_e\approx h$, and $\overline{w^\prime b^\prime}(h) \approx \overline{w^\prime b^\prime}(h_e)$.  Equating the empirical rule of convection and the boundary layer heat flux formula from \citet{Lilly1968} yields
\begin{equation}
\overline{w^\prime b^\prime}(h) = \Delta b \, \frac{\partial h}{\partial t} = 0.2 \, \overline{w^\prime b^\prime}_{\mbox{\tiny sfc}}.
\label{BL_buoyFlux}
\end{equation}
Using the assumed form of the entrainment layer buoyancy change, equation~(\ref{BL_buoyFlux}) becomes
\begin{equation}
\label{a:eqn}
a N^2 |h| \frac{\partial h}{\partial t} = a N^2 \frac{\partial}{\partial t}\left(\frac{h^2}{2}\right) = 0.2 \, \overline{w^\prime b^\prime}_{\mbox{\tiny sfc}}.
\end{equation}
Dividing equation~(\ref{a:eqn}) by equation~(\ref{apEq:3}) gives
\begin{equation}
\frac{a}{1-2a} = 0.2.
\label{aDef}
\end{equation}
Equation~(\ref{aDef}) gives $a\approx0.143$.  Inserting this result into equation~(\ref{apEq:3}), the final expression for boundary layer depth is
\begin{equation}
\label{finalH}
h(t) = \left(\frac{2.8 \; \overline{w^\prime b^\prime}_{sfc} \; t}{N^2}\right)^{1/2}.
\end{equation}

\clearpage

\bibliographystyle{agufull08}
\bibliography{CVMix_Papers.bib}

\end{document}